\documentclass[a4paper,11pt]{article}
\pdfoutput=1 

\usepackage{jheparxiv} 

\usepackage[T1]{fontenc} 

\usepackage{tensor}
\usepackage{amsmath}
\usepackage{amssymb}
\usepackage{mathrsfs}
\usepackage{graphicx}
\usepackage{stmaryrd}
\usepackage{twistor}
\usepackage{commath}
\usepackage{appendix}
\usepackage{yfonts}
\usepackage{tkz-euclide,subfigure}
\usepackage{capt-of}
\usepackage{xfp} 
\usepackage[outline]{contour} 
\usetikzlibrary{decorations.markings,decorations.pathmorphing}
\usetikzlibrary{angles,quotes} 
\usetikzlibrary{arrows.meta}
\usepackage{tikz-3dplot}
\usetikzlibrary{patterns}
\usetikzlibrary{math}
\contourlength{1.4pt}

\newcommand{\veps}{\varepsilon}

\newcommand{\msf}[1]{\mathsf{#1}}

\newcommand{\calI}{\mathscr{I}} 
\tikzset{>=latex} 
\colorlet{myred}{red!80!black}
\colorlet{myblue}{blue!80!black}
\colorlet{mygreen}{green!80!black}
\colorlet{mydarkred}{red!50!black}
\colorlet{mydarkblue}{blue!50!black}
\colorlet{mylightblue}{mydarkblue!6}
\colorlet{mypurple}{blue!40!red!80!black}
\colorlet{mydarkpurple}{blue!40!red!50!black}
\colorlet{mylightpurple}{mydarkpurple!80!red!6}
\tikzstyle{world line}=[myblue!60,line width=0.4]
\tikzstyle{world line t}=[mypurple!60,line width=0.4]
\tikzstyle{singularity}=[myred,line width=0.6,decorate,
                         decoration={zigzag,amplitude=2,segment length=6.17}]
\tikzset{declare function={%
  penrose(\x,\c)  = {\fpeval{2/pi*atan( (sqrt((1+tan(\x)^2)^2+4*\c*\c*tan(\x)^2)-1-tan(\x)^2) /(2*\c*tan(\x)^2) )}};%
  penroseu(\x,\t) = {\fpeval{atan(\x+\t)/pi+atan(\x-\t)/pi}};%
  penrosev(\x,\t) = {\fpeval{atan(\x+\t)/pi-atan(\x-\t)/pi}};%
  kruskal(\x,\c)  = {\fpeval{asin( \c*sin(2*\x) )*2/pi}};
}}

\def\Nsamples{40} 

\title{Celestial holography and $AdS_3$/$CFT_2$ from a scaling reduction of twistor space}

\abstract{Celestial amplitudes obtained from Mellin transforming 4d momentum space scattering amplitudes contain distributional delta functions, hindering the application of conventional CFT techniques. In this paper, we propose to bypass this problem by recognizing Mellin transforms as integral transforms projectivizing certain components of the angular momentum. It turns out that the Mellin transformed wavefunctions in the conformal primary basis can be regarded as representatives of certain cohomology classes on the minitwistor space of the hyperbolic slices of 4d Minkowski space. Geometrically, this amounts to treating 4d Minkowski space as the embedding space of $AdS_3$. By considering scattering of such on-shell wavefunctions on the projective spinor bundle $\mathbb{PS}$ of Euclidean $AdS_3$, we bypass the difficulty of the distributional properties of celestial correlators using the traditional $AdS_3/CFT_2$ dictionary and find conventional 2d CFT correlators for the scaling reduced Yang-Mills theory living on the hyperbolic slices. In the meantime, however, one is required to consider action functionals on the auxiliary space $\mathbb{PS}$, which introduces additional difficulties. Here we provide a framework to work on the projective spinor bundle of hyperbolic slices, obtained from a careful scaling reduction of the twistor space of 4d Minkowski spacetime.}

\author[a]{Wei Bu,}\emailAdd{w.bu@sms.ed.ac.uk} 
\author[b]{Sean Seet}\emailAdd{sxes2@cam.ac.uk} 

\affiliation[{a}]{School of Mathematics and Maxwell Institute for Mathematical Sciences\\
University of Edinburgh, EH9 3FD, UK}

\affiliation[b]{Department of Applied Maths \& Theoretical Physics,\\
Wilberforce Road, Cambridge CB3 0WA,\\
United Kingdom}

\begin{document} 
\maketitle
\flushbottom

\section{Introduction}
The celestial holography program proposes to look at 4d scattering amplitudes as 2d conformal correlators on the boundary celestial sphere \cite{Pasterski:2016qvg,Pasterski:2017kqt}. Celestial amplitudes are known to be harassed by distributional delta functions relating the holomorphic and antiholomorphic spinors, indicating unconventional features such as non-unitarity of the putative dual celestial CFT (CCFT). Different ways of regularizing such distributions include looking at celestial amplitudes in a different basis \cite{Sharma:2021gcz,Brown:2022miw,Hu:2022syq,De:2022gjn}, and considering theories coupled to an additional scalar profile breaking 4d Lorentz translation invariance \cite{Fan:2021pbp,Casali:2022fro}.

Inspired by the results in \cite{Casali:2022fro}, where celestial MHV amplitudes for a 4d Yang-Mills-Scalar theory were analyzed and the D-term \cite{DHoker:1999kzh} as in contact $AdS$ Witten diagrams were found with a particular choice of the scalar profile singling out a certain hyperbolic slice of Minkowski space. We look at the representation of pure 4d Yang-Mills theory on hyperbolic slices by scale reducing it to 3d Euclidean $AdS_3$. 

Many attempts have been made trying to relate flat space holography to the well-established $AdS/CFT$ \cite{deBoer:2003vf,Cheung:2016iub,Pasterski:2016qvg,Pasterski:2017kqt,Iacobacci:2022yjo,Sleight:2023ojm,Costello:2023hmi}. For example, \cite{Cheung:2016iub} focused on the leading soft sector of 4d Yang-Mills and performed its Kaluza-Klein reduction to hyperbolic slices of Minkowski space. A Chern-Simons/WZW model correspondence was found, capturing a subsector of the flat space hologram, with a completely chiral system on the boundary. This is very much analogous to \cite{Costello:2020jbh} where a chiral algebra on the celestial sphere was found capturing the dual 4d form factors in Yang-Mills theory. Only focusing on subsectors of the dynamics of the 4d theory clarifies the chiral sector of the dual 2d CFT, but further strategies are needed to describe the complete dual CCFT. Only recently, \cite{Costello:2022jpg,Costello:2023hmi} demonstrated the details of a genuine holographic duality between 4d WZW model coupled to Mabuchi gravity on the asymptotically flat Burns space and some large $N$ chiral algebra on the boundary celestial sphere. Certain quotients of Burns space can be identified with Euclidean $AdS_3$, it was mentioned in the paper that the Burns space holography can be regarded as compactification of the usual $AdS_3/CFT_2$ holography.

\medskip

In this paper, we shall see that viewing Mellin transform as a projective integral manifests the fact that Mellin transformed 4d Minkowski momentum eigenstates can also be naturally thought of as minitwistor representatives on the minitwistor space $\mathbb{MT}\cong(\mathbb{CP}^1\times\mathbb{CP}^1) \setminus \mathbb{CP}^1$ of Euclidean $AdS_3$ ($\mathbb{H}_3$) slices of 4d Minkowski space:
\begin{equation}
    \psi_M(\lambda,\mu)=\int_{\mathbb{R}_+} \frac{\d \omega}{\omega}\,\omega^{\Delta}\,\psi(\lambda,\omega\mu)\,,
\end{equation}
with $\lambda_\alpha,\mu^{\dal}$ coordinates on twistor space $\mathbb{PT}$ of Minkowski space. Explicit demonstration of this idea is presented in section \ref{section_mellin_as_proj}. Geometrically, this amounts to recognizing $\mathbb{R}^4$ as the embedding space of $\mathbb{H}_3$, where Mellin transforms on momentum eigenstates in $\mathbb{R}^4$ can be seen as dilatation eigenstates on $\mathbb{H}_3$ as in the embedding formalism.

To provide dynamical evidence of this observation, we obtain the MHV generating functional as a first order deformation away from the Bogomolny monopole sector on $\mathbb{H}_3$ using the scaling reduced twistor action of 4d Yang-Mills. We then evaluate the generating functional with on-shell states, which are the representatives obtained by Mellin transforming/projectivizing the 4d momentum eigenstates. This gives us the MHV amplitude on the hyperbolic slices, where the familiar D-term appears \cite{Simmons_Duffin_2014,DHoker:1999kzh}.

\medskip

The paper is organized as follows, section \ref{sec:2} quickly recaps the claim of celestial holography and the role of Mellin transform, together with the embedding formalism for correlators of CFTs living on $\mathbb{R}^d$ and their dual theories living on $\mathbb{H}_{d+1}$ in the usual $AdS/CFT$ holography. Section \ref{sec:3} reviews basics of twistor theory of 4d Minkowski space and highlights the Yang-Mills theory we are considering in 4d and review the twistor action used to compute its MHV amplitudes. Section \ref{section_minitwistor} introduces the geometry of minitwistor space and projective spinor bundle of 3d hyperbolic slices. Representatives of the cohomology classes on such auxiliary spaces are related to dynamical degree of freedoms on $\mathbb{H}_3$ through the Penrose transform. Section \ref{section_mellin_as_proj} then uses the geometry of these auxiliary spaces and introduces the main idea of the paper, regarding Mellin transform as giving independent projective scalings to twistor variables. By analysing the topology of the space, the Mellin transformed momentum space representatives can effectively be seen as representatives of cohomology classes on minitwistor space of $\mathbb{H}_3$. In section \ref{section_scaling_reduction} the actual scaling reduction of Yang-Mills is performed, an action functional for the scaling reduced theory is written down on the projective spinor bundle of $\mathbb{H}_3$. Its equivalence to the action on $\mathbb{H}_3$ is verified. The action is written as a perturbative expansion around the integrable Bogomolny subsector, section \ref{section_correlator} computes the MHV amplitudes using the MHV vertex part of the action where the D-term was found. In section \ref{discussion}, we finish with some discussions on the benefits of the formalism we set up compared to existing approaches to flat holography, and further ideas that could be easily implemented in this framework. More details on the embedding spaces and the twistor geometry are spelled out for interested readers in the appendices \ref{Appendix_A} and \ref{Appendix_B}.


\section{Backgrounds}\label{sec:2}
\subsection{Review of celestial holography} \label{section_bkgd}
Here we quickly recap the main idea of celestial holography introducing the role of Mellin transform, we refer the reader to more comprehensive and pedagogical reviews in the literature \cite{Pasterski:2021rjz,Raclariu:2021zjz}.
Celestial holography is a correspondence based on the isomorphism between the 4d Lorentz group and the 2d conformal group. The proposal entails regarding Mellin transformed massless momentum eigenstates in 4d Minkowski space $\mathbb{M}$ as 2d conformal primaries living on the celestial sphere at the asymptotic null boundary of $\mathbb{M}$. To explicitly see this, one uses spinor helicity variables to parametrize external null momenta $p_{\alpha\dal}$, such null momenta are parametrized by a point on the celestial sphere and a frequency/scale. Hence geometrically we should be able to use coordinates on this celestial sphere $z,\bar z$ to parametrize null momenta
\begin{equation}
    p_{\alpha\dal}= \kappa_{\alpha}\tilde\kappa_{\dal} = \omega \,\binom{1}{ z} (1\,,\bar z)\,,
\end{equation}
where we have selected a particular patch to parametrize the $SL(2,\mathbb{C})_R$ spinors, with $\omega$ being some energy scale. Mellin transform takes a momentum eigenstate and diagonalizes Lorentz boost along the direction of the null momenta:
\begin{equation}
    \phi_M(z,\bar z,\Delta) = \int_{\mathbb{R}^+}\frac{\d\omega}{\omega}\omega^{\Delta}\phi(p) = \int_{\mathbb{R}^+}\frac{\d\omega}{\omega}\omega^{\Delta}\e^{\im x^{\alpha\dal}\kappa_{\alpha}\tilde\kappa_{\dal}} = \frac{(-\im)^{-\Delta}\Gamma(\Delta)}{(x^{\alpha\dal}z_{\alpha}\bar z_{\dal})^{\Delta}}\,,
\end{equation}
where we have slightly abused notation by setting $z_{\alpha}=(1\,,z)$ and $\bar z_{\dal}=(1\,,\bar z)$. If one were to recognize $\Delta$ as labelling the conformal dimension of the Mellin transformed eigenstate inserted at position $(z,\bar z)$ on the celestial sphere, $\phi_M(z,\bar z,\Delta)$ looks like a conformal primary operator of a putative celestial CFT (CCFT) living on $S^2$. 

The dynamical aspect of the proposal then involves Mellin transforming usual scattering amplitudes in 4d Minkowski space and recasting them as conformal correlation functions of the CCFT:
\begin{equation}
    \prod_{i=1}^n\int\frac{\d\omega_i}{\omega_i}\,\omega_i^{\Delta_i} \mathcal{M}(\phi(p_1),\dots\phi(p_n)) = \mathcal{M}(\phi(z_1,\bar z_1,\Delta_1)\dots\phi(z_n,\bar z_n,\Delta_n))\,.
\end{equation}
The purpose of this paper is to point out the fact that it is natural to consider the Mellin transform as giving independent projective scaling to the combination $x_{\alpha\dal}\kappa^\alpha$, which then becomes a new momentum eigenstate on an auxiliary space associated to $\mathbb{H}_3$ called the minitwistor space. This allows one to use $AdS_3$/$CFT_2$ techniques to gain insights into the properties of the 2d CCFT living on the celestial sphere.

\begin{figure}[!]
    \centering
  \begin{tikzpicture}[scale=2.3]
  \def\Nlines{4} 
  \def\ta{tan(90*1.0/(\Nlines+1))} 
  \def\tb{tan(90*2.0/(\Nlines+1))} 
  \coordinate (O) at ( 0, 0); 
  \coordinate (S) at ( 0,-1); 
  \coordinate (N) at ( 0, 1); 
  \coordinate (W) at (-1, 0); 
  \coordinate (E) at ( 1, 0); 
  \fill[mylightblue] (N) -- (E) -- (S) -- (W) -- cycle;
  \node[mydarkblue,above right,align=left] at (55:0.7)
    {$\calI$};
  \node[mydarkblue,above left,align=right] at (125:0.7)
    {$\calI$};
  \foreach \i [evaluate={\c=\i/(\Nlines+1); \ct=tan(90*\c);}] in {1,...,\Nlines}{
    \message{  Running i/N=\i/\Nlines, c=\c, tan(90*\c)=\ct...^^J}
    \draw[world line t,samples=\Nsamples,smooth,variable=\t,domain=-1:1] 
      plot(\t,{-penrose(\t*pi/2,\ct)})
      plot(\t,{ penrose(\t*pi/2,\ct)});
  }
  \draw[thick,blue!60!black] (N) -- (E) -- (S) -- (W) -- cycle;
  \draw[thick,blue!60!black] (-1,0) -- (-2,-1);
  \draw[thick,blue!60!black] (-2,-1) -- (2,-1);
  \draw[thick,blue!60!black] (2,-1) -- (1,0);
  \draw[->,mydarkblue!80!black,shorten <=0.4] 
    (1,0) to[out=30,in=120]++ (-35:0.3)
    node[right] {\large $u=0$};
  \draw[->, thick] (0,-0.4) -- (0,0.4); 
  \node[left] at (0,0.1) {\Large $\tau$};
  \node[right] at (2.1,-1.02) {\large $i^0$};
  \node[above] at (0,1.02) {\large $i^+$};
\end{tikzpicture}
    \caption{Penrose diagram of the upper half of 4d Minkowski space, the shaded region is the focus of our consideration where hyperbolic slices are labelled by the coordinate $\tau$.}
    \label{figure1}
\end{figure}
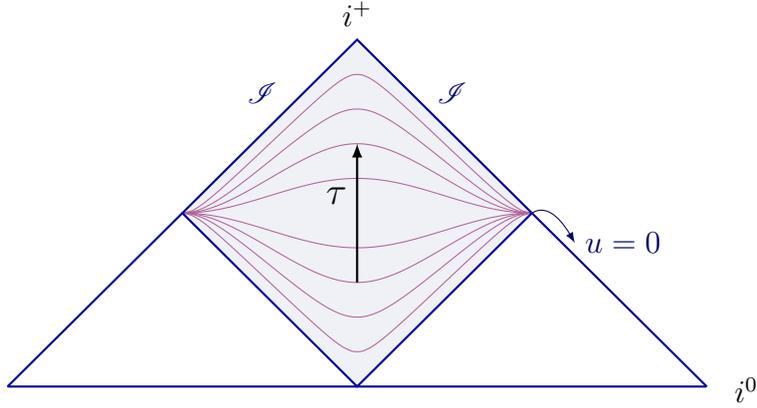

In order to make connection with the reinterpretation of the Mellin transform, we focus on particular $\mathbb{H}_3$ slices of complexified Minkowski space $\mathbb{M}_{\mathbb{C}}$\footnote{Note that when we work on the level of explicit amplitude computations, we take the dotted and undotted spinors to both be real and independent, which essentially amounts to picking the $(2,2)$ reality condition on $\mathbb{M}_{\mathbb{C}}$. This modifies the topology of the boundary celestial sphere from $S^2$ to $S^1\times S^1$, referred to as the celestial torus in the literature \cite{Atanasov:2021oyu,Mason:2022hly}.}. Each hyperbolic slice of the time-like region of $\mathbb{M}_{\mathbb{C}}$ is a copy of $\mathbb{H}_3$
\begin{equation}
    \d s^2 = -\d \tau^2 +\tau^2\d H_3^2\,,
\end{equation}
with $\tau^2=-x\cdot x$ the parameter labelling the hyperbolic slices. Note that as shown in figure \ref{figure1}, all the hyperbolic slices across different values of $\tau$ share the same boundary $S^2$, at retarded time $u=0$ on $\scri$.  

\subsection{Embedding formalism}
Here we assemble the facts and standardise the notation needed to use embedding space. There is a short introduction to the embedding space formalism in the appendix. Good references for the embedding space formalism for $\mathbb{R}^d$ include \cite{Weinberg_2010,Simmons_Duffin_2014}, and for $\mathbb{H}_{d+1}$ \cite{Costa_2014} and references therein. Consider Minkowski space:
\begin{equation}
x^{\alpha \dot \alpha} = \begin{pmatrix} t + z & x -i y \\ x + iy & t - z \end{pmatrix} \in \mathbb{R}^{3,1},
\end{equation}
 with the standard $\mathbb{R}^{3,1}$ metric: 
\begin{equation}
    \d s^2 = \d t^2 - \d x^2-\d y^2-\d z^2 = \epsilon_{\alpha \beta} \epsilon_{\dot \alpha \dot \beta} \d x^{\alpha \dot \alpha}\d x^{\beta \dot \beta}.
\end{equation}
We may embed $\mathbb{H}_{3}$ in $\{\mathbb{R}^{3,1}\setminus{0}\}$ by considering the projectivization: 
\begin{equation}
    x^{\alpha \dot \alpha} \sim r x^{\alpha \dot \alpha}\,,\,r \in \mathbb{R}^*\,.
\end{equation}
Restrict to the slice $x^2 > 0, x^{0\dot 0} \neq 0$ and consider the following metric:
\begin{equation}
    \d s^2_{\mathbb{H}_3} = \frac{1}{x^2}\left(\d x^2 - \frac{(x\cdot \d x)^2}{x^2}\right)\,.
\end{equation}
In the gauge in which $x^{0\dot 0} = 1$, we recover the Poincare half-space coordinates:
\begin{equation}
    x^{\alpha \dot \alpha} = \begin{pmatrix} 1 & x_1 -i x_2 \\ x_1 + ix_2 & x_0^2 + x_1^2+x_2^2 \end{pmatrix}=:\begin{pmatrix} 1 & \bar{z} \\ z & x_0^2 + z \bar z \end{pmatrix}\,.
\end{equation}
The conformal boundary of $\mathbb{H}_{3}$ is described in the embedding coordinates as the locus $x^2=0$. In the gauge in which $x^{0\dot 0} = 1$, this locus is described by:
\begin{equation}
    x^{\alpha \dot \alpha} = \begin{pmatrix} 1 & x_1 -i x_2 \\ x_1 + ix_2 & x_1^2+x_2^2 \end{pmatrix}=:\begin{pmatrix} 1 & \bar{z} \\ z & z \bar z \end{pmatrix}\,,
\end{equation}
where $x_1,x_2$ or $z, \bar{z}$ are coordinates on the boundary $S^2$. The notational advantage of working in embedding space is that conformal transformations on the boundary are realised as Lorentz transformations on the embedding space coordinates. In particular, the nonlinear realisation of the conformal generators can be seen to be the linear action of a Lorentz generator on the embedding space, and then a gauge transformation to return to the chosen gauge slice.

Note that the component Laplacian on $\mathbb{H}_3$ is represented in embedding space as:
\begin{equation}
    \square_{\mathbb{H}_3} = x^2 \partial_{\alpha \dot \alpha}\partial^{\alpha \dot \alpha} - \Upsilon(\Upsilon-2)\,,
\end{equation}
where $\Upsilon=x^{\alpha\dal}\partial_{\alpha\dal}$ is the Euler vector field. It is conventional that fields of defined dilatation weight are represented on the embedding space as fields of defined projective scaling weight, which we will abide by in order to make contact with the existing literature:
\begin{equation}
    \phi_{\Delta}(x_0,x_1,x_2) = \phi_{\Delta}\left(\frac{x^{\alpha \dot \alpha}}{|x|}\right) \rightarrow \Phi_{\Delta}(x^{\alpha \dot \alpha}) = (|x|)^\Delta \phi_{\Delta}\left(\frac{x^{\alpha \dot \alpha}}{|x|}\right)\,.
\end{equation}
It is important here to note the difference between dilatation weight and scaling weight. Although all components of $x^{\alpha \dal}$ have scaling weight 1 under $\Upsilon = x \cdot \partial/\partial x$, components of $x^{\alpha \dal}$ have different weights under dilatation, written in Poincare coordinates $(x_0,x_1,x_2)$ on the Poincare patch as:
\begin{equation}
    \hat \Delta x^{\alpha \dal} = \left(\sum_{i=0}^{2} x_i \frac{\partial}{\partial x_i}\right)\begin{pmatrix} 1 & x_1 -i x_2 \\ x_1 + ix_2 & x_0^2 + x_1^2+x_2^2 \end{pmatrix} = \begin{pmatrix} 0 & x_1 -i x_2 \\ x_1 + ix_2 & 2(x_0^2 + x_1^2+x_2^2) \end{pmatrix}\,.
\end{equation}
In particular, $x^2$ has dilatation weight = scaling weight = 2.
The prefactor $(|x|)^\Delta$ has dilatation weight to balance $\phi$, and has scaling weight $\Delta$. Hence multiplication sets the dilatation weight of $\Phi$ to 0 and sets the scaling weight of $\Phi$ to the dilatation weight of $\phi$. The procedure to recover the spacetime field is to multiply in the appropriate power of $|x|$ such that the expression is scale free, and then pick a gauge (we will frequently use $|x|=1$). Note that $|x|$ commutes with $\mathbb{H}_3$ derivatives, so multiplying it into the definition of fields does not change the field equations they satisfy. Finally, note that forms on the embedding space that descend to $\mathbb{H}_3$ must be basic under the Euler vector field. For a vector, this is:
\begin{equation}
    0 = i_{\Upsilon} A = A_{\alpha \dot \alpha} x^{\alpha \dot \alpha}\,,
\end{equation}
that is, they cannot "point" in the $\mathbb{R}^*$ fibre direction off of $\mathbb{H}_3$ into the ambient space. A basis decomposition that conveniently encodes this condition is:
\begin{equation}
    A_{\alpha \dot \alpha} \d x^{\alpha \dot \alpha} = \frac{1}{x^2}A_{\dot \alpha (\alpha} x^{\dot \alpha}_{\beta)} (x^{(\alpha |\dot \gamma|} \d x_{\dot \gamma}^{\beta)}) =: A_{\alpha \beta} \frac{\D x^{\alpha \beta}}{x^2}\,,
\end{equation}
where $A_{\alpha \beta} = A_{\dot \alpha (\alpha} x^{\dot \alpha}_{\beta)}$. The analogous statement for spinors $\Psi$ is $X^\mu \Gamma_{4 \mu} \Psi = 0$, but will play no role in this paper.

\section{Twistor theory in 4d and the Yang-Mills twistor action}\label{sec:3}
\subsection{Twistor theory in 4d}\label{section_4d_twistor}
Twistor space $\mathbb{PT}$ of 4d complexified Minkowski space $\mathbb{M}_{\mathbb{C}}$ is defined as an open subset of the collection of all the complex lines through $\mathbb{C}^4$, $\mathbb{PT}:=\mathbb{CP}^3\setminus\mathbb{CP}^1$. If we coordinatize $\mathbb{C}^4$ with $Z^{I}= (\lambda_{\alpha}, \mu^{\dal})$, where $\alpha=0,1$, $\dal=\dot 0, \dot 1$ are $SL(2,\mathbb{C})$ indices. Then $\mathbb{CP}^3$ can be defined as the identification $Z^I$ with $sZ^I$ with nonzero complex number $s$ and $\mathbb{PT}$ is given by the open subset of $\mathbb{CP}^3$ where $\lambda_{\alpha}\neq 0$. $\mathbb{PT}$ and the original complexified Minkowski space coordinatized by $x^{\alpha\dal}$ are related by a geometric relation called the incidence relation 
\begin{equation}\label{incidence_relation_r4}
    \mu^{\dal} = x^{\alpha\dal}\lambda_{\alpha}\,.
\end{equation}
Geometrically, this means that each point $x\in\mathbb{M}_{\mathbb{C}}$ corresponds to a complex projective line $L_x\cong\mathbb{CP}^1$ linearly embedded in $\mathbb{PT}$. To see this, we temporarily unprojectivize $\mathbb{PT}$, \eqref{incidence_relation_r4} defines a $\mathbb{C}^2$ with fixed $x^{\alpha\dal}$. Restoring the projective scaling we find $\mathbb{CP}^1\subset\mathbb{CP}^3$ \cite{Adamo:2017qyl}. Alternatively, each fibre of the fibration of $\mathbb{PT}$ over $\mathbb{M}_{\mathbb{C}}$ is a projective line $\mathbb{CP}^1$. 

Conversely, taking two such projective lines corresponding to two distinct points $x$ and $y$ in $\mathbb{M}_{\mathbb{C}}$ incident at a point, \eqref{incidence_relation_r4} gives 
\begin{equation}
    (x-y)^{\alpha\dal} = \lambda^{\alpha}\tilde\iota^{\dal}\,,
\end{equation}
for some arbitrary spinor $\tilde\iota^{\dal}$. Varying $\tilde\iota^{\dot 0}$ and $\tilde\iota^{\dot 1}$ sweeps out null complex 2-plane called the $\alpha$-plane $\mathbb{C}^2\subset\mathbb{M}_{\mathbb{C}}$. Hence a point in $\mathbb{PT}$ corresponds to an $\alpha$-plane in $\mathbb{M}_{\mathbb{C}}$.

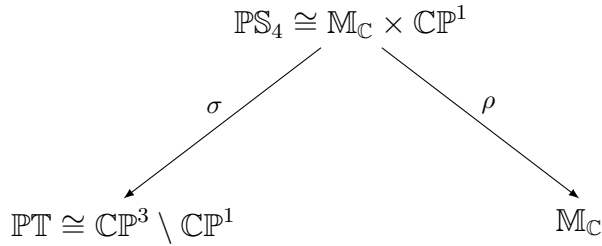
\begin{figure}
    \centering
    \begin{tikzpicture}
\draw[->] (2.6,2) --  (0,0);
\draw[->] (3.4,2) --  (6,0);
\tkzDefPoint(0,0){A}
\tkzDefPoint(6,0){B}
\tkzDefPoint(3,2){C}
\tkzDefPoint(4.8,1){D}
\tkzDefPoint(1.2,1){E}
\tkzLabelPoint[below](A){\large $\mathbb{PT}\cong\mathbb{CP}^3\setminus\mathbb{CP}^1$}
\tkzLabelPoint[below](B){\large $\mathbb{M}_\mathbb{C}$}
\tkzLabelPoint[above](C){\large $\mathbb{PS}_4\cong\mathbb{M}_\mathbb{C}\times\mathbb{CP}^1$}
\tkzLabelPoint[above](D){\small $\rho$}
\tkzLabelPoint[above](E){\small $\sigma$}
\end{tikzpicture}
    \caption{Double fibration of the projective spinor bundle over $\mathbb{PT}$ and $\mathbb{R}^4$ with $\rho$ the obvious forgetful map and $\sigma$ the incidence relation.}
    \label{figure_PT}
\end{figure}
The intersection of this $\alpha$-plane with the real Euclidean slice of $\mathbb{R}^4 \in \mathbb{M}_{\mathbb{C}}$ must be dimension $0$ as there are no null directions in Euclidean $\mathbb{R}^4$. By linearity, the intersection is either empty or a single point. We can explicitly solve for the point $x\in\mathbb{R}^4$ and show that the intersection is nonempty:
\begin{equation}
   x^{\alpha \dot \alpha} = \hat x^{\alpha \dot \alpha} \implies x^{\alpha \dot \alpha} = \frac{\hat \mu^{\dot \alpha}\lambda^{\alpha}-\mu^{\dot \alpha}\hat \lambda^{\alpha}}{\langle \lambda \hat \lambda \rangle}\,,
\end{equation}
where we define the following involution on dotted and undotted spinors:
\begin{align}
    &l^{\alpha} = \left(l^{0},l^{1}\right)\mapsto \hat l^{\alpha} = \left(-\bar l^{1},\bar l^{0}\right)\,,\\
    &m^{\dot \alpha} = \left(m^{\dot 0},m^{\dot 1}\right)\mapsto \hat m^{\dot \alpha} = \left(-\bar m^{\dot 1},\bar m^{\dot 0}\right)\,.
\end{align}
such that the real slice picked out of $\mathbb{C}^4$ coordinatised by $x^{\alpha \dot \alpha} = \hat x^{\alpha \dot \alpha}$ is real Euclidean:
\begin{align}
    x^{\alpha \dot \alpha} =  \hat x^{\alpha \dot \alpha} &\implies x^{\alpha \dot \alpha} = \begin{pmatrix}
        it+z & x-iy \\ x+iy & it-z
    \end{pmatrix}, \quad (t,x,y,z) \in \mathbb{R}^4,
    \\ \textrm{such that } &\implies \epsilon_{\alpha \beta}\epsilon_{\dot\alpha \dot\beta}x^{\alpha \dot \alpha}x^{\beta \dot \beta} = -t^2-x^2-y^2-z^2\,.
\end{align}
In this way, a point in $\mathbb{PT}$ (that corresponds to an alpha plane in $\mathbb{M_C}$) determines a point in $\mathbb{R}^4$. Therefore there exists a projection map $\pi: \mathbb{PT} \rightarrow \mathbb{R}^4, Z^A \rightarrow x^{\alpha \dot \alpha}$ with a 2-real-dimensional kernel, and it is natural to consider $\mathbb{PT}$ as the total space of a fibre bundle over $\mathbb{R}^4$ with a 2-real-dimensional fibre. The conventional description is what is known as the projective spin bundle, $\mathbb{PS}_4$, coordinatized by $(\lambda_{\alpha},x^{\alpha \dot \alpha}) \in \mathbb{PS}_4 \cong\mathbb{CP}^1 \times\mathbb{R}^4 $, the trivial bundle of projectivized undotted Weyl spinors over $\mathbb{R}^4$. As a real manifold, $\mathbb{PS}_4 \cong \mathbb{PT}$, with $(\lambda_{\alpha}, x^{\alpha \dot \alpha}\lambda_{\alpha})=(\lambda_{\alpha}, \mu^{\dot \alpha})$. 
\begin{figure}
    \centering
    \begin{tikzpicture}
\draw[->] (2.6,2) --  (2.6,0);
\tkzDefPoint(2.6,2.1){A}
\tkzDefPoint(2.6,-0.1){B}
\tkzLabelPoint[above](A){\large $\mathbb{PT}\cong\mathbb{PS}_4\cong\mathbb{R}^4\times\mathbb{CP}^1$}
\tkzLabelPoint[below](B){\large $\mathbb{R}_4$}
\end{tikzpicture}
    \caption{Fibration of Euclidean $\mathbb{PT}$ over $\mathbb{R}^4$.}
    \label{figure_PT2}
\end{figure}

\medskip

The power of such non-local construction is that cohomology classes on $\mathbb{PT}$ are in one-to-one correspondence with linearized zero-rest-mass fields on Minkowski space through the Penrose transform \cite{Penrose:1985bww,Penrose:1986ca}. For example, a scalar representative is an element of the cohomology class living on $\mathbb{PT}$ 
\begin{equation}
    \psi(Z^I) \in H^{0,1}(\mathbb{PT},\mathcal{O}(-2))\,,
\end{equation}
meaning it is a $\bar \partial$-closed $(0,1)$-form which scales as $\psi(sZ^I)=s^{-2}\psi(Z^I)$. How do we obtain some object that purely depends on the spacetime coordinate $x^{\alpha\dal}$? The answer lies in the double fibration diagram as in figure \ref{figure_PT}. Starting with objects on $\mathbb{PT}$, we would like to use the map $\sigma$ to pull the representative $\psi(\lambda_{\alpha},\mu^{\dal})$ back to $\mathbb{PS}$, where we replace all the $\mu^{\dal}$ with $x^{\alpha\dal}\lambda_{\alpha}$ using the incidence relation. Geometrically this means we are using \eqref{incidence_relation_r4} to restrict $\psi$ to the projective line $L_x$ corresponding to a point $x\in\mathbb{M}$. Then the $\rho$ map taking us down to $\mathbb{R}^4$ just asks for an integration over different choices of $\lambda_{\alpha}$s. The composition of these two maps is the Penrose integral formula 
\begin{equation}
    \Psi(x) = \int_{\mathbb{CP}^1}\D\lambda \left.\psi(\lambda,\mu)\right\vert_{L_x}\,,
\end{equation}
where $\Psi(x)$ automatically satisfies the zero rest mass equation $\Box\Psi = 0$ as a consequence of holomorphicity.

\subsection{Twistor action for 4D Yang-Mills}\label{section_6dYM}
In this section we review the twistor action for 4d Yang-Mills theory \cite{Mason:2005zm,Boels:2006ir} and the basic MHV amplitude computation through holomorphic frames \cite{Boels:2007qn,Bu:2022dis,Bullimore:2011ni,Adamo:2011cb}. It is a well-known fact that 4d Yang-Mills theory can be written as a perturbative expansion around an integrable subsector, self-dual Yang-Mills. 
\begin{multline}
    S[A] = -\frac{1}{2g^2} \int_{\mathbb{R}^4} \tr\left(F^{(4)}\wedge \star F^{(4)}\right) \simeq -\frac{1}{2g^2} \int_{\mathbb{R}^4} \tr\left(F^{-}\wedge F^{-}\right) \\
    \stackrel{\text{EOM of G}}{=} \int_{\mathbb{R}^4} \tr \left(F^-\wedge G\right)+\frac{g^2}{2}\int_{\mathbb{R}^4}\tr\left( G\wedge G\right)\label{YM_action_spacetime}\,,
\end{multline}
where the $\simeq$ sign means they are equal considered up to a topological term and the last equality holds after integrating out $G$ in the path integral. $F^-$ is the anti-self-dual part of the curvature and $G$ is some anti-self-dual 2-form taking values in $\mathfrak{g}$. Through the Ward correspondence \cite{WARD197781,Witten:2003nn}, the self-dual part of the action can be straightforwardly lifted to twistor space
\begin{equation}\label{S_sdym_6d}
    S_{SDYM} = \int_{\mathbb{PT}}\D^3Z\wedge\tr\left( B\wedge \left(\bar\partial A+A\wedge A\right)\right)\,,
\end{equation}
with $B\in \Omega^{0,1}(\mathbb{PT}, \mathcal{O}(-4)\otimes \mathfrak{g})$ and $A\in \Omega^{0,1}(\mathbb{PT}, \mathfrak{g})$ the negative and positive helicity gluon representatives on twistor space. The equation of $B$ sets the antiholomorphic part of the curvature to zero, by the Ward correspondence \cite{WARD197781}, a holomorphic gauge bundle on twistor space is equivalent to self-duality equation on $\mathbb{R}^4$. The full Yang-Mills action is obtained by adding a non-local term
\begin{equation}\label{S_int}
    S_{\text{int}} = \int_{\mathbb{R}^4} \d^4x \int_{\mathbb{CP}^1\times\mathbb{CP}^1} \D\lambda_1\D\lambda_2\la\lambda_1\lambda_2\ra^2\, \tr\left(B(x,\lambda_1)W_{12} B(x,\lambda_2)W_{21}\right) \,,
\end{equation}
where $W_{12}$ and $W_{21}$ are holomorphic Wilson lines parallel transporting the gauge bundle between different points on $\mathbb{CP}^1$. 
\begin{equation}
    W_{z_1z_2} =1+\sum_{m=1}^{\infty}(-1)^m\int_{(\mathbb{CP}^1)^m}\frac{(z_2-z_1)}{(z_2-\sigma_m)(\sigma_m-\sigma_{m-1})\cdots(\sigma_1-z_1)}\bigwedge_{i=1}^m A_i(\sigma_i)\d\sigma_i\,,
\end{equation}
where $\sigma_i$s are coordinates on $\mathbb{CP}^1$. It has been shown that with appropriate gauge choices, $S_{\text{int}}$ reduces to $\int_{\mathbb{R}^4}\tr\left( G\wedge G\right)$ on $\mathbb{R}^4$. The full Yang-Mills twistor action is then:
\begin{equation}\label{YM_6d}
    S_{YM}= S_{SDYM}+S_{\text{int}}\,.
\end{equation}
MHV amplitude with helicity configuration $--+++\dots$ can be obtained by evaluating $S_{\text{int}}$ using on-shell momentum eigenstates
\begin{equation}
    A^{\msf{a}} =t^{\msf{a}} \int_{\mathbb{C}^*_s} \frac{\d s}{s} \bar\delta^2(\kappa_{\alpha}-s\lambda_{\alpha})\,\e^{\im s\omega[\mu\bar z]} \quad,\quad B^{\msf{a}}= t^{\msf{a}} \int_{\mathbb{C}^*_s} \d s s^3\bar\delta^2(\kappa_{\alpha}-s\lambda_{\alpha})\,\e^{\im s\omega[\mu\bar z]}\,.
\end{equation}
Substituting into \eqref{S_int}, notice there is a holomorphic delta function for each integral over $\mathbb{CP}^1$. $\eqref{S_int}$ evaluates to the familiar Parke-Taylor factor \cite{Parke:1986gb}
\begin{equation}
    \mathcal{M}_{MHV_n} = \tr(t^{\msf{a}_1}...t^{\msf{a}_n})\frac{\la\kappa_1\kappa_2\ra^4}{\la\kappa_1\kappa_2\ra\cdots\la\kappa_{n-1}\kappa_{n}\ra\la\kappa_n\kappa_1\ra}\,\int_{\mathbb{R}^4}\d^4 x \e^{\im \left(x \cdot \sum_{i=1}^np_i\right)} \,.
\end{equation}


\section{Twistor theory in 3d}\label{section_minitwistor}

Here we provide a very swift recap of the required twistor theory in 3d (minitwistor theory). More details of the discussion are added in appendix \ref{appendix_minitwistor} based on material in the literature \cite{Bailey_Dunne_1998,Huggett_Tod_1985,Penrose:1986ca,Jones:1985pla}. A good discussion of the projective spinor bundle $\mathbb{PS}$ of $\mathbb{R}^3$ can be found in \cite{Adamo:2017xaf}.

\subsection{$\mathbb{H}_3$ Minitwistors and the projective spin bundle $\mathbb{PS}$}\label{subsection_minitwistor1}

\begin{center}
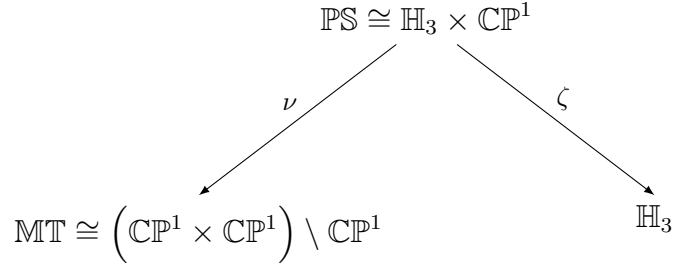

\begin{tikzpicture}
\draw[->] (2.6,2) --  (0,0);
\draw[->] (3.4,2) --  (6,0);
\tkzDefPoint(0,0){A}
\tkzDefPoint(6,0){B}
\tkzDefPoint(3,2){C}
\tkzDefPoint(4.8,1){D}
\tkzDefPoint(1.2,1){E}
\tkzLabelPoint[below](A){\large $\mathbb{MT}\cong\left(\mathbb{CP}^1\times\mathbb{CP}^1\right)\setminus\mathbb{CP}^1$}
\tkzLabelPoint[below](B){\large $\mathbb{H}_3$}
\tkzLabelPoint[above](C){\large $\mathbb{PS}\cong\mathbb{H}_3\times\mathbb{CP}^1$}
\tkzLabelPoint[above](D){\small $\zeta$}
\tkzLabelPoint[above](E){\small $\nu$}
\end{tikzpicture}
\captionof{figure}{Double fibration of $\mathbb{PS}$ over $\mathbb{MT}$ and hyperbolic three space with $\zeta$ the obvious forgetful map and $\nu$ the incidence relation.} 
\end{center}

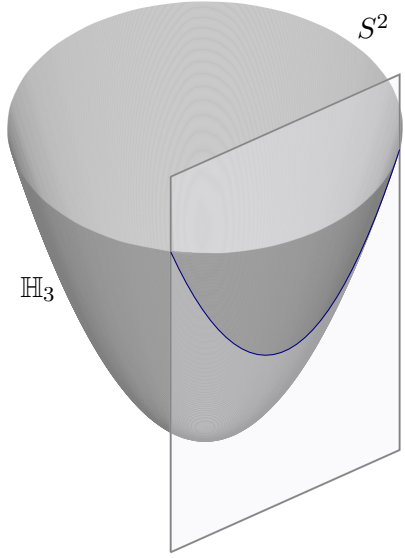
\begin{figure}
    \centering
    \tdplotsetmaincoords{50}{100}
\begin{tikzpicture}[tdplot_main_coords,scale=1.3]
		\tikzmath{function f(\x) {return \x;};}
		\pgfmathsetmacro{\zini}{0.5*sqrt(2.0)}
		\pgfmathsetmacro{\step}{0.01}
		\pgfmathsetmacro{\zsig}{\zini+\step}
		\pgfmathsetmacro{\nextz}{\zini+0.5*\step}
		\pgfmathsetmacro{\sig}{2.0*\step}
		\pgfmathsetmacro{\tini}{0.5*pi}
		\pgfmathsetmacro{\tfin}{1.85*pi}
		\pgfmathsetmacro{\tend}{2.5*pi}

		\foreach \altura in {\step,\sig,...,4.0}{
			\pgfmathparse{sqrt(\altura)}
			\pgfmathsetmacro{\radio}{\pgfmathresult}
			\draw[gray,opacity=0.42] plot[domain=0:6.2832,smooth,variable=\t] ({\radio*cos(\t r)},{\radio*sin(\t r)},{\altura}); 
		}
		\fill[color=mylightblue,opacity=0.3] (0,2,0) -- plot[domain=0:2, samples=30] (\x,2-\x,{2*(\x)^2 - 4*(\x) + 4})  -- (2,0,0) -- (0,2,0);
  		\fill[color=mylightblue,opacity=0.3] (0,2,5) -- plot[domain=0:2, samples=30] (\x,{(4-(\x)^2)^(1/2)},4)  -- (2,0,5) -- (0,2,5);
		\draw[thick,gray!70] (2,0,0) -- (2,0,5) -- (0,2,5) -- (0,2,0) -- (2,0,0);
        \draw[thick,gray!70] plot[domain=0:6.2832,smooth,variable=\t] ({2*cos(\t )},{2*sin(\t)},4);
		\draw[gray] (2,0,0) -- (2,0,5);
		\draw[gray] (0,2,0) -- (0,2,5);
        \draw[gray] (2,0,5) -- (0,2,5);
        \draw[gray] (2,0,0) -- (0,2,0);
		\node[left] at (1,-1.25,2.5) {$\mathbb{H}_3$};	
        \node[above] at (2.1,2.1,7.1) {$S^2$};
        \draw[mydarkblue] plot[domain=0:2, samples=30] (\x,2-\x,{2*(\x)^2 - 4*(\x) + 4});
\end{tikzpicture}
    \caption{Here we draw the geometric picture for $\mathbb{H}_3$ in $\mathbb{R}^{3,1}$ embedding space. The plane contains the origin and a null geodesic (that does not contain the origin) in the embedding space. The intersection of the plane and a hyperboloid is a geodesic on $\mathbb{H}_3$ connecting two points on the conformal boundary $S^2$.} 
    \label{figure3}
\end{figure}

The minitwistor space of a spacetime is the space of oriented geodesics on the spacetime. The minitwistor space of $\mathbb{H}_3$ is denoted $\mathbb{MT}\cong (\mathbb{CP}^1 \times \mathbb{CP}^1) \setminus \mathbb{CP}^1$:
\begin{equation}
    \mathbb{MT} = \{(\lambda_{\alpha},\mu^{\dot\alpha}) \in \mathbb{CP}^1 \times \mathbb{CP}^1 | \la \bar \mu \lambda \ra \neq 0\};
\end{equation}
with incidence relation:
\begin{equation}
 \mu^{\dot\alpha} = x^{\alpha \dot\alpha} \lambda_{\alpha}\,,
\end{equation}
where $x^{\alpha \dot \alpha} \in \mathbb{CP}^3$ is a point in $\mathbb{H}_{3 \mathbb{C}}$ via the embedding space formalism, and we introduce the Lorentzian involution:
\begin{align}
    &l^{\alpha} = \left(l^{0},l^{1}\right)\mapsto \bar l^{\dot \alpha} = \left(\bar l^{0},\bar l^{1}\right)\,,\\
   & m^{\dot \alpha} = \left(m^{\dot 0},m^{\dot 1}\right)\mapsto \bar m^{\alpha} = \left(\bar m^{\dot 0},\bar m^{\dot 1}\right)\,.
\end{align}
For a given point in $\mathbb{MT}$, the locus of the incidence relation describes the geodesic on $\mathbb{H}_{3}$ that interpolates between two points on the conformal boundary $S^2$. More explicitly, the two boundary points can be written in Lorentzian embedding space coordinates as:
\begin{equation}
    z^{\alpha \dot\alpha} = \bar \mu^{\alpha} \mu^{\dot\alpha} \quad ,\quad w^{\alpha \dot\alpha} = \lambda^{\alpha} \bar \lambda^{\dot \alpha}\,.
\end{equation}
We can then parametrize points $x^{\alpha\dal}\in AdS_3$ using these two boundary points up to scale:
\begin{equation}
    x^{\alpha \dot \alpha} \sim  z^{\alpha \dot \alpha} + tw^{\alpha \dot \alpha}\,.
\end{equation}
As shown in figure \ref{figure3}, varying $t$ traces out the blue line which is the geodesics in $AdS_3$. 

The removal of the antiholomorphic diagonal $\mathbb{CP}^1$ where $\la\bar\mu \lambda\ra = 0$ corresponds to the removal of the region where $z=w$ (up to projective rescalings), in which the geodesic degenerates into a single boundary point. The corresponding picture in the 4D embedding space is that the null 2-plane touches the boundary $S^2$ at a point. On restriction to the real Lorentzian slice of $\mathbb{M_C}$, this means that the 4d real Lorentzian null geodesic (defined by the intersection of the null 2-plane with the real Lorentzian slice) is contained in the 4d null cone of the origin.

Note that $\overline{\la\bar\mu \lambda\ra}=[\mu\bar\lambda]=x_{\alpha\dal}\lambda^{\alpha}\bar\lambda^{\dal}=0$ is where the boundary $S^2$ of $\mathbb{H}_3$ lies in 4d Minkowski space coordinate. This matches the geometric picture we started with, as depicted in figure \ref{figure1}, actually all the $\mathbb{H}_3$s we consider have a common boundary $S^2$ at $u=0$. The minitwistor space can be found by a complex scaling reduction of the flat 4D twistor space $\mathbb{PT}$ discussed in section \ref{section_4d_twistor}, and then the removal of the antiholomorphic diagonal. The construction is summarised briefly in the appendix.
\begin{equation}
    \mathbb{MT} = \{ \mathbb{PT} / \mu^{\dot\alpha} \sim r\mu^{\dot\alpha}\,,\,r\in\mathbb{C}^*\} \setminus \{\la\bar\mu\lambda\ra = 0\}\,.
\end{equation}
Note that the minitwistor space does not fibre over the spacetime - regardless of reality condition, a point in minitwistor space corresponds to a geodesic, which is an extended object in $\mathbb{H}_{3\mathbb{C}}$. A closely related space that does fibre over the spacetime for particular reality conditions can be found by a real scaling reduction on the flat 4D twistor space. We denote the space $\mathbb{PS}$:
\begin{equation}
    \mathbb{PS} = \{ \mathbb{PT} / \mu^{\dot\alpha} \sim r\mu^{\dot\alpha}\,,\,r\in\mathbb{R}^*\}\,.
\end{equation}
With incidence relation:
\begin{equation}
 \mu^{\dot\alpha} = x^{\alpha \dot\alpha} \lambda_{\alpha} \,,
\end{equation}
where $x^{\alpha \dot \alpha} \in \mathbb{C}^4 / 
\mathbb{R}^*$. Under real Euclidean reality conditions, the locus of the incidence relations for a fixed $(\lambda,\mu)$ is a point in the real Euclidean embedding space description of $S^3$, where we will analytically continue the results of calculations to $\mathbb{H}_3$.

\subsection{Penrose Transform and Hitchin-Ward correspondence}\label{section_penrose_transform_mt}
Recall that we define $f \in H^{0,1}(\mathbb{MT},\mathcal{O}(-p-2+\Delta, -\Delta))$ by picking out the eigenstate of $\mu$ scaling with eigenvalue $-\Delta$ from some $F \in H^{0,1}(\mathbb{PT},\mathcal{O}(-p-2))$:
\begin{equation}
    f = \int_{\mathbb{R}^+} \frac{\d\omega}{\omega} e^{ln(\omega)(\mu\cdot \partial_\mu + \Delta)} F(\lambda,\mu) = \int_{\mathbb{R}^+} \frac{d\omega}{\omega} \omega^{\Delta} F(\lambda,\omega \mu )\,.
\end{equation}
The machinery of embedding space allows us to import standard flat 4D twistor theory results to $\mathbb{MT}$. The standard Penrose Transform for $p \geq 0$ descends to $\mathbb{MT}$ \cite{Baston:1989vh,Tsai1996ThePT}:
\begin{align}
    \{f\in H^{0,1}(\mathbb{MT},\mathcal{O}(-p-2+\Delta, -\Delta))\} \cong \{\ \phi_{\alpha_1 ... \alpha_p}(x) \nonumber
    \\ | \square_{\mathbb{H}_3} \phi_{\alpha_1 ... \alpha_p}(x) = \Delta(\Delta+2) \phi_{\alpha_1 ... \alpha_p}, \quad \partial^{\alpha_1 \dot\alpha} \phi_{\alpha_1 ... \alpha_p}(x) = 0\}\,.
\end{align}
This is implied by the Penrose transform relating $F\in H^{0,1}(\mathbb{PT},\mathcal{O}(-p-2))$ and solutions to the linearized zero rest mass equations on the embedding space. Similarly, the Penrose-Ward correspondence (\cite{WARD197781,Hitchin:1982gh,Hitchin:1983ay} and \cite{Tsai1996ThePT}) can be written down on $\mathbb{MT}$. Here we present the integral form of the transform with $a \in H^{0,1}(\mathbb{MT},\mathcal{O}(\Delta, -\Delta))$ as an example:
\begin{equation}
    w_{\alpha\dal}= |x|^{\Delta}\int_{\mathbb{CP}^1} \D\lambda\,\frac{\iota_{\alpha}}{\la\iota\lambda\ra}\,\left.\frac{\partial a}{\partial\mu^{\dal}}\right\vert_{L_x}\,, 
\end{equation}
where $\iota_{\alpha}$ is some reference spinor and $\left.\right\vert_{L_x}$ denotes pulling back the cohomology class to the $\mathbb{CP}^1$ fibre. Such integral transforms are called the Sparling transforms, their equivalence to the usual Penrose transforms are demonstrated in appendix \ref{appendix_sparling_transform}. In the embedding space, $w$ has a self dual (linearised) field strength. 
\begin{equation}
    F_{\alpha \beta} = \partial^{\dot \alpha}_{(\alpha} w_{\beta) \dot \alpha} = 0\,,
\end{equation}
The analogous statement on $\mathbb{H}_3$ is that $w$ encodes both a gauge field as well as a scalar field 
\begin{equation}
    \Phi=w\cdot x  \,,\quad A_{\alpha\dal}=w_{\alpha\dal}-\frac{w\cdot x}{x^2}\,x_{\alpha\dal} \,,
\end{equation}
and that they satisfy the linearized Bogomolny equation \cite{Atiyah:1988jp}:
\begin{equation}
    \star\d A =  \d\Phi\,,
\end{equation}
We touch more upon the relationship between this equation and the self-duality equation in 4d in section \ref{section_YMH3}. Hence actual particles in the scattering processes we will consider are combinations of $A_{\alpha\dal}$ and $\Phi$.

\section{Mellin transform as a projective integral}\label{section_mellin_as_proj}
Equipped with the understanding of the geometry of twistor space $\mathbb{PT}$ and minitwistor space $\mathbb{MT}$ of $\mathbb{H}_3$, we are now ready to present the main observation of the paper. Recall that in section \ref{section_4d_twistor}, homogeneous coordinates $\mu^{\dal}$ and $\lambda_{\alpha}$ have the same complex projective scalings as coordinates on $\mathbb{PT}$. Taking a representative $\psi(\lambda,\mu)\in H^{0,1}(\mathbb{PT},\mathcal{O}(-2))$, we can give $\mu$ independent scalings through the following Mellin transform:
\begin{equation}\label{Mellin_transform}
    \psi_M(\lambda,\mu)=\int_{\mathbb{R}_+} \frac{\d \omega}{\omega}\,\omega^{\Delta}\,\psi(\lambda,\omega\mu)\,.
\end{equation}
Instead of looking at $\psi_M$ as an eigenstate in the conformal primary basis following the usual proposal of celestial holography, here the purpose of the Mellin transform is to give the twistor coordinate $\mu^{\dal}$ independent scaling weights from $\lambda_{\alpha}$. We shall see what benefits this perspective on Mellin transform brings us. As an explicit example, we evaluate such a projective integral \eqref{Mellin_transform} by picking some momentum eigenstate on $\mathbb{PT}$, which is a particular representative of $H^{0,1}(\mathbb{PT},\mathcal{O}(-2))$:
\begin{align}
   \psi_M(\lambda,\mu)&= \int_{\mathbb{R}_+} \frac{\d \omega}{\omega}\,\omega^{\Delta}\,\psi(\lambda,\omega\mu) \nonumber\\
   &= \int_{\mathbb{R}_+} \frac{\d \omega}{\omega}\,\omega^{\Delta}\,\int_{\mathbb{C}^*_s} \d s s \bar\delta^2(\kappa_{\alpha}-s\lambda_{\alpha})\,\e^{\im s\omega[\mu\bar z]} = \frac{(-\im)^{-\Delta}\bar\delta(\la\lambda\kappa\ra)\la\lambda\iota\ra^{\Delta-1}\Gamma(\Delta)}{\la\kappa\iota\ra^{\Delta-1}[\mu\bar z]^{\Delta}}\label{Momentum_eigenstate}\,,
\end{align}
where $\iota^{\alpha}$ is a constant reference spinor which can be chosen arbitrarily. Note that we have only parametrized the dotted spinor on a local patch $\tilde\kappa^{\dal}=\omega(1\,,\bar z)$ while keeping the undotted spinors in global coordinates. Notice that the original representative $\psi(q\lambda,q\mu)$ scales as $q^{-2}\psi(\lambda,\mu)$, which is what was meant by a $(0,1)$-form twisted by line bundle $\mathcal{O}(-2)$. Whereas the Mellin transformed eigenstate \eqref{Momentum_eigenstate} can be seen as having independent projective scalings on the two twistors, $\psi_M(q\lambda,p\mu)\mapsto q^{\Delta-2}p^{-\Delta}\psi_M(\lambda,\mu)$, indicating the scaling weights of $\psi_M(\lambda,\mu)$ to be $(\Delta-2,-\Delta)$.
Penrose transform of $\psi_M$ to spacetime gives
\begin{equation}
    \int_{\mathbb{CP}^1}\D\lambda \left.\frac{(-\im)^{-\Delta}\bar\delta(\la\lambda\kappa\ra)\la\kappa\iota\ra^{\Delta+1}}{\la\lambda\iota\ra^{\Delta+1}[\mu\bar z]^{\Delta}}\right\vert_{\mu=x\lambda}=\frac{(-\im)^{-\Delta}}{(x^{\alpha\dal}\kappa_{\alpha}\tilde\kappa_{\dal})^{\Delta}}= \frac{(-\im)^{-\Delta}}{(x\cdot p)^{\Delta}}\,,
\end{equation}
which we recognize as the $\mathbb{H}_3$ bulk-to-boundary propagator for a scalar conformal primary \cite{Costa:2014kfa}. As the reader might have guessed, the appearance of $\mathbb{H}_3$ here is not a coincidence, this perspective on the Mellin transformed eigenstate also changed the topology of the space we are looking at. The $\psi_M(\lambda,\mu)$ with independent complex projective scalings on its coordinates suggests that the space we are looking at is no longer $\mathbb{PT}\cong\mathbb{CP}^3\setminus\mathbb{CP}^1$, but rather $\mathbb{MT}\cong(\mathbb{CP}^1\times\mathbb{CP}^1)\setminus\mathbb{CP}^1$. $\mathbb{MT}$ is the minitwistor space of $\mathbb{H}_3$ we introduced in \ref{subsection_minitwistor1}. Together with the scaling weights we observed earlier, the Mellin transformed representative $\psi_M$ in \eqref{Momentum_eigenstate} can be seen as a representative of $H^{0,1}(\mathbb{MT},\mathcal{O}(\Delta-2,-\Delta))$, whose spacetime counterpart can be worked out using the Sparling transform reviewed in section \ref{section_penrose_transform_mt}.


As mentioned in section \ref{section_bkgd}, Minkowski space admits hyperbolic foliations with each leaf being a copy of $\mathbb{H}_3$. In order for this perspective on Mellin transform to shed some light on celestial holography, we choose to demonstrate our proposal of understanding celestial holography by computing MHV amplitudes in scaling reduced Yang-Mills theory living on the hyperbolic slices. Using the Yang-Mills twistor actions reviewed in section \ref{section_6dYM}, we shall perform the scaling reduction to arrive at an action on minitwistor space for the scaling reduced Yang-Mills theory living on hyperbolic slices labelled by constant $\tau$ in section \ref{section_scaling_reduction}. 
\section{Scaling reduction}\label{section_scaling_reduction}
As promised in section \ref{section_mellin_as_proj}, we would like to examine MHV amplitudes in the representative of 4d Yang-Mills on hyperbolic slices, obtained through scaling reduction along $\tau$. Before diving into the reduction of the twistor action \eqref{YM_6d}, we first describe the scaling reduced Yang-Mills action on $\mathbb{H}_3$, and its relation to the usual 3d Yang-Mills-Higgs theory on $\mathbb{H}_3$. 
\subsection{Scaling reduced Yang-Mills on $\mathbb{H}_3$}\label{section_YMH3}
One could start with Yang-Mills theory in 4d flat space
\begin{equation}
    S[A^{(4)}] = -\frac{1}{2g^2} \int_{\mathbb{R}^3\times S^1} \tr(F^{(4)}\wedge \star F^{(4)})\,,
\end{equation}
and scaling reduce by introducing a Higgs field $\Phi$ to obtain Yang-Mills-Higgs theory on 3d flat space 
\begin{equation}
    S[A,\Phi] = -\frac{1}{2g^2}\int_{\mathbb{R}^3} \tr(F^{(3)}\wedge \star F^{(3)}+ \D_A\Phi \wedge \star \D_A\Phi)\,,
\end{equation}
where the 4d theory is reduced on a circle of vanishing radius. $D_A$ here refers to the standard covariant derivative. As mentioned in section \ref{section_6dYM}, 4d Yang-Mills theory can be written as a perturbative expansion around the self-dual instanton sector \cite{WARD197781,Chalmers:1996rq}, which is integrable and admits a twistor description. As described in detail in \cite{Adamo:2017xaf}, the subsector relation is inherited in $\mathbb{R}^3$, the 3d Yang-Mills-Higgs theory can be perturbatively expanded around the Bogomolny monopole sector $\star F = \D\Phi$, which similarly is also integrable and admits a minitwistor description \cite{Hitchin:1983ay}. 
\begin{equation}
     S[A,\Phi]\simeq\int_{\mathbb{R}^3} \tr[B\wedge (F-\star \D_A\Phi)] +\frac{g^2}{2}\int_{\mathbb{R}^3}\tr(B \wedge \star B)\,,
\end{equation}
where again the $\simeq$ sign suggests that they differ by a topological term $\frac{1}{g^2}\int_{\mathbb{R}^3}\tr\left(F\wedge\D_A\Phi\right)$. Such topological terms vanish as long as the manifold has no boundary. However, in the case of $\mathbb{H}_3$, the boundary does exist so one simply cannot write Yang-Mills-Higgs as a perturbative expansion around the Bogomolny subsector. Hence the theory we are looking at here is not exactly Yang-Mills-Higgs in $\mathbb{H}_3$, it is only perturbatively equivalent to that with a difference of a topological term.
Hence the action on $\mathbb{H}_3$ obtained from scale reducing Yang-Mills as a perturbation around the self-dual sector \eqref{YM_action_spacetime} along $\tau$:
\begin{equation}\label{YMH_ads3}
    S[A,B,\Phi]_{\mathbb{H}_3} = \int_{\mathbb{H}_3} \tr[B\wedge (F-\star \D_A\Phi)] +\frac{g^2}{2}\int_{\mathbb{H}_3}\tr(B \wedge \star B)
\end{equation}
is not equivalent to the usual Yang-Mills-Higgs theory in 3d:
\begin{equation}
    S[A,\Phi]_{\mathbb{H}_3} =  -\frac{1}{2g^2}\int_{\mathbb{H}^3} \tr(F^{(3)}\wedge \star F^{(3)}+ \D_A\Phi \wedge \star \D_A\Phi)\,.
\end{equation}
They differ by a topological term\footnote{We thank Tim Adamo for discussion of this point}:
\begin{equation}
    \int_{\mathbb{H}_3} \tr\left(\D_A(\Phi\wedge  F)\right) = \int_{S^2} \tr\left(\Phi\wedge F\right)\,.
\end{equation}
We shall refer to the theory \eqref{YMH_ads3} as scaling reduced Yang-Mills (srYM) to manifest the difference with Yang-Mills-Higgs. For our purpose in this paper, it is indifferent to us whether the scaling reduced theory is Yang-Mills-Higgs or not, as long as it comes from scaling reduction of 4d Yang-Mills on the hyperbolic slices. However, if one were to be interested in actual Yang-Mills-Higgs theory in $\mathbb{H}_3$, one could add such BF action on the boundary $S^2$ to compensate for the perturbative difference.

\subsection{Twistor action}
The discussion on scale reducing 4d Yang-Mills to scale reduced Yang-Mills in $\mathbb{H}_3$ in section \ref{section_YMH3} extends to their counterparts on twistor space. Scaling reduction of the Yang-Mills twistor action \eqref{YM_6d} to minitwistor space is actually rather interesting\footnote{The reduction to $\mathbb{MT}$ of $\mathbb{H}_3$ we are trying to do is actually the complementary of the $S^3$ reduction performed in \cite{Costello:2020jbh}, where we would like to reduce on the radial direction of $S^3\subset\mathbb{R}^4$ rather than keeping it.}. As mentioned in section \ref{section_bkgd}, we shall work with the coordinate
$\tau^2=-x^{\alpha\dal}x_{\alpha\dal}=\frac{2[\hat{\mu}\mu]}{\la\lambda\hat{\lambda}\ra}$ which labels the hyperbolic slices, scale reducing along $\mathbb{C}^*_{\tau}$ should give us an action on minitwistor space. Working in Euclidean signature on twistor space of $\mathbb{R}^4$, we take a basis of antiholomorphic forms adapted to the dimensional reduction. Adapting the usual differentials on $\mathbb{PT}$ to accommodate the scaling along $\mathbb{C}^*_{\tau}$:
\begin{equation}
    \bar\partial = \frac{\D\hat{\lambda}}{\la\hat{\lambda}\lambda\ra}\lambda_{\alpha}\frac{\partial}{\partial\hat{\lambda}_{\alpha}}+\frac{[\hat\mu \d\hat\mu]}{[\mu\hat\mu]}\mu^{\dal}\frac{\partial}{\partial\hat\mu^{\dal}}+\frac{2\bar\partial \tau}{\tau}\hat\mu^{\dal}\frac{\partial}{\partial\hat\mu^{\dal}}\,.
\end{equation}
We then decompose the fields in \eqref{YM_6d} using the dual basis 
\begin{equation}
    A= a_\lambda\frac{\D\hat{\lambda}}{\la\lambda\hat{\lambda}\ra^2} +a_{\tau}\bar\partial\tau+a_{\mu}\frac{[\hat{\mu}\d\hat{\mu}]}{\tau^4\la\lambda\hat{\lambda}\ra^2}\,, \quad\quad    B= b_\lambda\frac{\D\hat{\lambda}}{\la\lambda\hat{\lambda}\ra^2} +b_{\tau}\bar\partial\tau+b_{\mu}\frac{[\hat{\mu}\d\hat{\mu}]}{\tau^4\la\lambda\hat{\lambda}\ra^2}\,.
\end{equation}
First we perform the scaling reduction of the self-dual part of the action \eqref{S_sdym_6d}, the holomorphic part of the measure $\D^3Z$ has the following decomposition in terms of the basis we just described:
\begin{equation}
    D^3Z\propto \D\lambda\wedge\partial \tau\wedge[\mu\d\mu]\,.
\end{equation}
Collecting all the terms remaining in the scaling reduction, we have the following for the BF part of the minitwistor action:
\begin{equation}
    \int_{\mathbb{MT}}\D\lambda\wedge[\mu\d\mu]\wedge\tr\left(\eta\wedge F(a)+b\wedge \bar\partial_a\xi\right)\,,
\end{equation}
where $\bar\partial_a=\bar\partial+a$ is the covariant antiholomorphic derivative and $a_\tau$ was relabeled as $\xi$ and $b_\tau$ as $\eta$. $b\in \Omega^{0,1}(\mathbb{MT},\mathcal{O}(\Delta-3,-\Delta-1))$ and $a\in \Omega^{0,1}(\mathbb{MT},\mathcal{O}(\Delta-1,-\Delta+1))$ are the Mellin transformed representatives on minitwistor space. We have kept all the  dynamical information in the twistor space eigenstates by considering arbitrary values of $\Delta$. We also note the scaling weights of the three $a$ wavefunctions in $F(a)=\bar\partial a+a\wedge a$ are different, chosen as long as the entire integrand has zero scaling weight. 

As noted in section \ref{section_minitwistor}, $\mathbb{MT}$ is not directly fibred over $\mathbb{H}_3$, therefore it is more convenient to work with $\mathbb{PS}\cong \mathbb{H}_3\times\mathbb{CP}^1$ if our purpose here is to see the equivalence to the scaling reduced Yang-Mills on $\mathbb{H}_3$, which can simply be achieved by integrating over the $\mathbb{CP}^1$ fibre. Hence we promote both $\eta$ and $\xi$ to be real 1-forms pointing along the resurrected $\tau'$ direction on $\mathbb{PS}$, and relax the scaling reduction $\mathbb{PS}$. In practice, this means that the scaling reduction is performed not on $\mathbb{C}^*_{\tau}$, but rather on $\mathbb{R}^*_{\tau'}$, where
\begin{equation}
    d\tau' =\frac{2\left([\hat{\mu}\d\mu]-[\mu\d\hat{\mu}]\right)}{\la\lambda\hat{\lambda}\ra}\,.
\end{equation}
Such a differential mixes the original holomorphic and antiholomorphic forms degrees, making the resulting projective spinor bundle a complex-real (CR) manifold. Hence the $(2,3)$-form action functional is written on the CR manifold $\mathbb{PS}$:
\begin{equation}\label{Bogomolny_PS_action}
    \int_{\mathbb{PS}} \D\lambda\wedge[\mu\d\mu]\wedge\tr\left(\eta\wedge F(a)+b\wedge \bar\partial_a\xi +a\wedge \d b\right)\,,
\end{equation}
where $\bar\partial_a$ is the antiholomorphic covariant derivative pointing in $\mu$ and $\lambda$ direction, $\d$ is the real derivative pointing in the $\mathbb{R}$ direction. Penrose transform for the on-shell fields works as described in section \ref{section_penrose_transform_mt}, further detail can be found in \cite{Tsai1996ThePT}. 

The reduction of the MHV vertex \eqref{S_int} is also straightforward, the measures in the spacetime direction nicely recombine into $\D^3 x$ with the vertex pointing along $\mathbb{CP}^1_{\lambda}$ direction unaffected:
\begin{equation}\label{MHVvertex}
    \int_{\mathbb{H}_3}\frac{\D^3 x}{x^4}\int_{\mathbb{CP}^1\times\mathbb{CP}^1} \D\lambda_1\D\lambda_2\la\lambda_1\lambda_2\ra^2\,\tr\left( b(\lambda_1)W_{12} b(\lambda_2)W_{21} \right)\,.
\end{equation}
Since the original 6d $B$s only pointed along the $\mathbb{CP}^1_{\lambda}$ direction, which was left invariant under the scaling reduction along $\tau'$. Combining \eqref{Bogomolny_PS_action} and \eqref{MHVvertex}, we have the scaling reduced Yang-Mills minitwistor action on $\mathbb{PS}$
\begin{equation}\label{PS_full_action}
    S_{\mathbb{PS}} = \int_{\mathbb{PS}}\D\lambda\wedge\D\mu\wedge \mathbf{b}\wedge F(\mathbf{a})+  \int_{\mathbb{H}_3}\frac{\D^3 x}{x^4}\int_{\mathbb{CP}^1\times\mathbb{CP}^1} \D\lambda_1\D\lambda_2\la\lambda_1\lambda_2\ra^3\, \tr\left(b(\lambda_1)W_{12} b(\lambda_2)W_{21} \right)\,,
\end{equation}
where we have abbreviated the terms in \eqref{Bogomolny_PS_action} using $\mathbf{a}$ and $\mathbf{b}$ to include the recently promoted $\xi$ and $\eta$:
\begin{equation}
    \mathbf{a} = \xi+a  \quad \text{and}\quad \mathbf{b} = \eta+b\,.
\end{equation}
We shall see in the next section that it is equivalent to the action on $\mathbb{H}_3$ \eqref{YMH_ads3} after gauge fixing, in particular, as we have stressed in section \ref{section_YMH3}, it differs from actual Yang-Mills-Higgs theory in $\mathbb{H}_3$ by a topological term. 

\subsection{Equivalence on spacetime}
We begin by reducing the Bogomolny monopole part of the minitwistor action \eqref{Bogomolny_PS_action} to $\mathbb{H}_3$ to find the Bogomolny sector of the spacetime action. First we specify the basis and dual vectors. The 1-forms that descend from the antiholomorphic basis in the flat 4d twistor space are:
\begin{equation}
    \bar e_{\mu} = 2\frac{\hat\lambda^{\alpha}\hat\lambda^{\beta}(x\d x)_{\alpha\beta}}{x^2\la\lambda\hat\lambda\ra^2}\,, \quad\quad e_{\phi} = -4\frac{\lambda^{(\alpha}\hat{\lambda}^{\beta)}(x\d x)_{\alpha\beta}}{x^2\la\lambda\hat{\lambda}\ra} \,,\quad\quad \bar e_{\lambda} = \frac{\D\hat{\lambda}}{\la\lambda\hat{\lambda}\ra^2}\,,
\end{equation}
where $(x\d x)_{\alpha\beta}=\veps^{\dal\Dot{\beta}}x_{\alpha\dal}\d x_{\beta\Dot{\beta}}$. The direction we have got rid of in the scaling reduction was $e_\tau=2\frac{\lambda^{[\alpha}\hat{\lambda}^{\beta]}(x\d x)_{\alpha\beta}}{x^2\la\lambda\hat{\lambda}\ra}=\frac{xdx}{x^2}$. The vectors dual to these are
\begin{equation} 
    \bar\partial_{\mu} = x_{\beta}{}^{\dal}\lambda^{\alpha}\lambda^{\beta}\frac{\partial}{\partial x^{\alpha\dal}}\,,\quad\quad \d_{\phi} = x_{\beta}{}^{\dal}\frac{\lambda^{(\alpha}\hat{\lambda}^{\beta)}}{\la\lambda\hat{\lambda}\ra}\,\frac{\partial}{\partial x^{\alpha\dal}}\,, \quad\quad \bar\partial_{\lambda} =\langle \lambda \hat \lambda \rangle \lambda^{\alpha}\,\frac{\partial}{\partial\hat{\lambda}^{\alpha}}\,.
\end{equation}
Abusing notations, we assemble them into a $\bar \partial$ operator on $\mathbb{PS}$ that is to be understood as descending from the flat 4D twistor space $\bar \partial$:
\begin{equation}
    \bar \partial = \bar e_{\mu} \bar \partial_{\mu} + e_{\phi} d_{\phi} + \bar e_{\lambda} \bar \partial _{\lambda}\,.
\end{equation}
The 1-forms that descend from the holomorphic basis in the flat 4D twistor space are 
\begin{equation}
    e_{\mu} = 2\frac{\lambda^{\alpha}\lambda^{\beta}(x\d x)_{\alpha\beta}}{x^2} \,,\quad\quad e_{\lambda} = \D\lambda\,.
\end{equation}
This choice of basis is appealing as it is very clear how we have decomposed the 4 components of $(x^{\alpha \dot \alpha} \d x_{\dot \alpha}^{\beta})$. The downside of working in this basis is the frame dragging, as the $\bar \partial_{\lambda}$ operator acts as:
\begin{equation}
    \bar \partial_\lambda (\bar e_{\mu}, e_{\phi} ) = (e_{\phi}, -2 e_\mu)\,.
\end{equation}

We collect all the measures in \eqref{Bogomolny_PS_action}:
\begin{equation}
    \D\lambda\wedge\D\mu\wedge \bar e_{\mu}\wedge\bar e_{\lambda}\wedge e_{\phi} = -4 \frac{\D\lambda\D\hat{\lambda}}{\la\lambda\hat{\lambda}\ra^2}\wedge\frac{\D^3x}{x^4}\,,
\end{equation}
where $\D^3x=(x\d x)^{\alpha\beta}(\d x\d x)_{\alpha\beta}$. 
Expanding the fields in components:
\begin{align}
    & a = a_\mu \bar e_{\mu}+a_{\lambda} \bar e_{\lambda} +a_{\phi} e_{\phi}\nonumber\\
    & b = b_\mu \bar e_{\mu}+b_{\lambda} \bar e_{\lambda} +b_{\phi} e_{\phi}\,.
\end{align}
Note that due to the frame dragging, we have, for instance:
\begin{multline}
    \bar \partial a = (\bar\partial_{\mu}a_{\phi}-\d_{\phi}a_{\mu}) \bar e_{\mu} \wedge e_{\phi} + (\bar\partial_{\lambda}a_{\phi} - d_{\phi} a_{\lambda}+ a_{\mu})\bar e_{\lambda} \wedge e_{\phi}+ \nonumber
    \\(\bar\partial_{\lambda}a_{\mu} - \bar\partial_{\mu}a_{\lambda}) \bar e_{\lambda} \wedge \bar e_{\mu} -2 a_{\phi} \bar e_{\lambda} \wedge e_{\mu}\,,
\end{multline}
where the two terms with no derivatives come from the frame dragging on the basis forms. Imposing harmonic gauge on $\mathbb{CP}^1_{\lambda}$ sets $a_{\lambda}=0$ and $b_{\lambda}=\frac{B_{\alpha\beta}\hat{\lambda}^{\alpha}\hat{\lambda}^{\beta}}{\la\lambda\hat{\lambda}\ra^2}$ for some spacetime field $B_{\alpha \beta}(x) = B_{(\alpha \beta)}(x)$ \cite{Woodhouse:1985id}, which kills some terms in the action \eqref{Bogomolny_PS_action} and leaves us with
\begin{equation}
    -4\int_{\mathbb{PS}}\frac{\D\lambda\wedge\D\hat{\lambda}}{\la\lambda\hat{\lambda}\ra^2}\wedge\frac{\D^3x}{x^4}\,\left(b_{\mu}\left(\bar\partial_{\lambda}a_{\phi}+ a_{\mu}\right)-b_{\phi}\bar\partial_{\lambda}a_{\mu}+b_{\lambda}(\d_{\phi}a_{\mu}-\bar\partial_{\mu}a_{\phi}+[a_{\phi},a_{\mu}])\right)\,.
\end{equation}
Equation of motions of $b_{\mu}$ and $b_{\phi}$ set: 
\begin{equation}
    \bar\partial_{\lambda}a_{\phi}+ a_{\mu}=0 \,,\quad\quad \bar\partial_{\lambda}a_{\mu} = 0\,.
\end{equation}
Given their respective weights $0$ and $2$, we have for some spacetime fields $\Phi(x)$, and $A_{\alpha\beta}(x)=A_{(\alpha\beta)}(x)$
\begin{equation}
    a_{\phi}=-A_{\alpha\beta}\frac{\lambda^{\alpha}\hat \lambda^{\beta}}{\langle \lambda \hat \lambda \rangle} + \Phi \,,\quad\quad a_{\mu} = A_{\alpha\beta}\lambda^{\alpha}\lambda^{\beta}\,.
\end{equation}
Substituting the fields with explicit $\lambda$ dependence in the remaining terms into the action 
\begin{equation}
    \int_{\mathbb{PS}}\frac{\D\lambda\wedge\D\hat{\lambda}}{\la\lambda\hat{\lambda}\ra^2}\wedge\frac{\D^3x}{x^4}\,\left(b_{\lambda}(\d_{\phi}a_{\mu}-\bar\partial_{\mu}a_{\phi}+[a_{\phi},a_{\mu}])\right)\,,
\end{equation}
we have the following
\begin{multline}
    -4\int_{\mathbb{H}_3\times\mathbb{CP}^1}\frac{\D\lambda\wedge\D\hat{\lambda}}{\la\lambda\hat{\lambda}\ra^2}\wedge\frac{\D^3x}{x^4}\times\\
    \frac{B_{\alpha\beta}\hat{\lambda}^{\alpha}\hat{\lambda}^{\beta}}{\la\lambda\hat{\lambda}\ra^2}\,\left(x_{\gamma}{}^{\dal}\frac{\lambda^{\delta}\hat\lambda^{\gamma}}{\la\lambda\hat{\lambda}\ra}\frac{\partial}{\partial x^{\delta\dal}}\,A_{\eta\epsilon} \lambda^{\eta}\lambda^{\epsilon} - x_{\kappa}{}^{\Dot{\beta}}\lambda^{\kappa}\lambda^{\rho}\frac{\partial}{\partial x^{\rho\Dot{\beta}}}\left(-A_{\zeta\theta}\frac{\lambda^{\zeta}\hat \lambda^{\theta}}{\langle \lambda \hat \lambda \rangle} + \Phi\right)\right.+
    \\\left. \left[\left(-A_{\chi\upsilon}\frac{\lambda^{\chi}\hat \lambda^{\upsilon}}{\langle \lambda \hat \lambda \rangle} + \Phi\right),A_{\tau\omega}\lambda^{\tau}\lambda^{\omega}\right] \right)\,.
\end{multline}
Since the fields $A_{\alpha\beta}$ and $\Phi$ only depend on $\mathbb{H}_3$ coordinates $x$, we can evaluate the $\mathbb{CP}^1$ integral explicitly (as a consequence of Serre duality on $\mathbb{CP}^1$):
\begin{align}
    &\int \frac{\D \lambda \wedge \D \hat \lambda }{\la \lambda \hat \lambda \ra^2} \frac{\lambda^{\alpha_1}...\lambda^{\alpha_n} \hat \lambda_{\beta_1}...\hat \lambda_{\beta_n}}{\la \lambda \hat \lambda \ra^n} = \int \D \lambda \wedge \D \hat \lambda \frac{(-1)^n}{(n+1)!} \lambda^{\alpha_1}...\lambda^{\alpha_n} \frac{\partial}{\partial \lambda^{\beta_1}} ... \frac{\partial}{\partial \lambda^{\beta_n}}\frac{1}{\la \lambda \hat \lambda \ra^2} \nonumber
    \\= &\int \frac{\D \lambda \wedge \D \hat \lambda}{\la\lambda \hat\lambda \ra^2} \frac{1}{n+1} \delta^{(\alpha_1}_{(\beta_1} ... \delta^{\alpha_n)}_{\beta_n)} = \int \d\theta \int \frac{r \d r}{(1+r^2)^2}\frac{1}{n+1} \delta^{(\alpha_1}_{(\beta_1} ... \delta^{\alpha_n)}_{\beta_n)} \nonumber
    \\= &\frac{2 \pi }{n+1} \delta^{(\alpha_1}_{(\beta_1} ... \delta^{\alpha_n)}_{\beta_n)}\,.
\end{align}
The integral that we will have to do for reductions to spacetime is the following:
\begin{align}
    \int \frac{\D \lambda \wedge \D \hat \lambda }{\la \lambda \hat \lambda \ra^2} \frac{\lambda^{\alpha_1}\lambda^{\alpha_2}\lambda^{(\alpha_3} \hat \lambda^{\beta_1)}\hat \lambda^{\beta_2}\hat \lambda^{\beta_3}}{\la \lambda \hat \lambda \ra^n} =  \frac{2 \pi }{4} \delta^{(\alpha_1}_{(\gamma_1}\delta^{\alpha_2}_{\gamma_2} \delta^{\alpha_3}_{\gamma_3)}\epsilon^{\beta_1) \gamma_3}\epsilon^{\beta_2 \gamma_2}\epsilon^{\beta_3 \gamma_1}\,.
\end{align}
Performing the integral over $\mathbb{CP}^1$ in each term, we have:
\begin{align}
    -2\pi \int_{\mathbb{H}_3} \frac{\D^3 x}{x^4} B_{\alpha \beta} \left((x\partial)^{\beta \gamma} A^{\alpha}_{\gamma} - [A^{\alpha \gamma}, A_{\gamma}^{\beta}] +(x \partial)^{\alpha \beta}\Phi -[A^{\alpha \beta},\Phi]\right)
    \\
    =-2\pi \int_{\mathbb{H}_3} B \wedge (\D A - \star \D \Phi)\label{reduction_middle_BF}\,,
\end{align}
where we use the basic embedding space 1-form $\D x^{\alpha \beta} = (x\d x)^{(\alpha \beta)}$
\begin{equation}
    B = B_{\alpha \beta} \frac{\D x^{\alpha \beta}}{x^2}, \quad A = A_{\alpha \beta} \frac{\D x^{\alpha \beta}}{x^2}, \quad \star \, \frac{\D x^{\alpha \beta}}{x^2} = \frac{(\d x \d x)^{\alpha \beta}}{x^2} = 2 \frac{\D x^{\alpha \gamma} \wedge \D x^{\beta}_{\gamma}}{x^4}\,.
\end{equation}
The reduction of the non-local part of the action \eqref{MHVvertex} is straightforward as the $a_{\lambda}$ component vanishes in harmonic gauge, hence the holomorphic Wilson lines reduce to $1$. The $b$s in the action are in fact $b_{\lambda}\bar e_{\lambda}$s, summarizing all the ingredients we write down \eqref{MHVvertex} in harmonic gauge:
\begin{equation}
    \int_{\mathbb{H}_3}\frac{\D^3 x}{x^4} \int_{\mathbb{CP}^1\times\mathbb{CP}^1} \frac{\D\lambda_1\wedge\D\hat{\lambda}_1}{\la\lambda_1\hat{\lambda}_1\ra^2}\wedge\frac{\D\lambda_2\wedge\D\hat{\lambda}_2}{\la\lambda_2\hat{\lambda}_2\ra^2}\, \frac{B_{\alpha\beta}(x)\hat{\lambda}_1^{\alpha}\hat{\lambda}_1^{\beta}}{\la\lambda_1\hat{\lambda}_1\ra^2}\, \frac{B_{\gamma\delta}(x)\hat{\lambda}_2^{\gamma}\hat{\lambda}_2^{\delta}}{\la\lambda_2\hat{\lambda}_2\ra^2}\la\lambda_1\lambda_2\ra^3\,.
\end{equation}
Doing the $\mathbb{CP}^1$ integrals gives us 
\begin{equation}
    \int_{\mathbb{H}_3}\tr\left(B\wedge \star B\right)\,.
\end{equation}
Together with the BF-part of the action \eqref{reduction_middle_BF}, the action we obtain on spacetime is
\begin{equation}
    S[A,B,\Phi] = \int_{\mathbb{H}_3} \tr[B\wedge (F-\star D\Phi)] +\frac{g^2}{2}\int_{\mathbb{H}_3}\tr(B \wedge \star B)\,,
\end{equation}
which matches the action on $\mathbb{H}_3$ \eqref{YMH_ads3}.

\section{The correlator}\label{section_correlator}
Now equipped with the minitwistor action of scaling reduced Yang-Mills theory \eqref{PS_full_action}, we would like to look at MHV scattering of the representatives on $\mathbb{PS}$ pulled back from $\mathbb{MT}\cong\mathbb{CP}^1\times\mathbb{CP}^1$, which we saw in section \ref{section_mellin_as_proj}, amounts to taking the Mellin transform of the usual 4d momentum eigenstate
\begin{equation}\label{PS_reps}
    a = \Gamma(\Delta-1)\frac{(-\im)^{1-\Delta}\bar\delta(\la\lambda\kappa\ra)}{[\mu\bar z]^{\Delta-1}}\,\left(\frac{\la\iota\lambda\ra}{\la\iota\kappa\ra}\right)^{\Delta}\,; \quad\quad b =\Gamma(\Delta+1)\frac{(-\im)^{-1-\Delta}\bar\delta(\la\lambda\kappa\ra)}{[\mu\bar z]^{\Delta+1}}\,\left(\frac{\la\iota\lambda\ra}{\la\iota\kappa\ra}\right)^{\Delta-2}\,.
\end{equation}
Their corresponding Penrose transform give
\begin{equation}
    |x|^{\Delta-1} \int\D\lambda \frac{\hat\lambda_{\alpha}}{\la \lambda \hat \lambda \ra} \frac{\partial}{\partial \mu^{\dot\alpha}} \Gamma(\Delta-1)\frac{(-\im)^{1-\Delta}\bar\delta(\la\lambda\kappa\ra)}{[\mu\bar z]^{\Delta-1}}\,\left(\frac{\la\iota\lambda\ra}{\la\iota\kappa\ra}\right)^{\Delta} = |x|^{\Delta}\,\frac{\hat\kappa_{\alpha}\bar z_{\dot \alpha}}{\la\kappa\hat\kappa\ra}\,\frac{(-\im)^{1-\Delta}\Gamma(\Delta-1)}{(x\cdot(\kappa\bar z))^{\Delta-1}}\,,
\end{equation}
similarly for $b$. As previously noted in section 5.2, this is a combination of the bulk-to-boundary propagators for the scalar $\phi$ and the gauge field $A$. The expression is self-dual when pulled back to embedding space, and Bogomolny on $\mathbb{H}_3$. We recover the gauge field by projecting out the scaling direction, and a short computation shows that the gauge field component $A_{\alpha \beta} = A_{\dot\alpha (\alpha} x^{\dot \alpha}_{\beta)}$ is:
\begin{equation}
    A_{\alpha \beta} = |x|^{\Delta-1}\,\frac{x^{\dot \alpha}_{(\beta}\hat\kappa_{\alpha)}\bar z_{\dot \alpha}}{\la\kappa\hat\kappa\ra}\,\frac{(-\im)^{1-\Delta}\Gamma(\Delta-1)}{(x\cdot(\kappa\bar z))^{\Delta-1}}\,.
\end{equation}
The MHV vertex in the minitwistor action \eqref{MHVvertex} functions in a similar fashion as \eqref{S_int} in the twistor action in section \ref{section_6dYM}.
We could bring down as many holomorphic frames containing $a$ as desired from the holomorphic Wilson line in \eqref{MHVvertex}, which gives
\begin{multline}
    \int_{\mathbb{H}_3}\frac{\D^3 x}{x^4}\int_{L_x}\D\lambda_1 \,|x|^{\Delta_1+1}\,\tr\left(b(\lambda_1)\left(\prod_{i=2}^{n-1}\frac{\D\lambda_i}{\la\lambda_i\lambda_{i+1}\ra}\, |x|^{\Delta_i-1}\,a(\lambda_{i+1})\right)\,\right.\\
   \left. \frac{\D\lambda_n}{\la\lambda_{n-1}\lambda_n\ra} \,|x|^{\Delta_n+1}\,b(\lambda_n)\la\lambda_1\lambda_n\ra^2\right)\,.
\end{multline}
Substituting in the eigenstates \eqref{PS_reps}, we have
\begin{multline}\label{n_correlator}
    \widetilde{\mathcal{M}}_{MHV_n}=\frac{\la\kappa_1\kappa_n\ra^4}{\la\kappa_1\kappa_2\ra\cdots\la\kappa_{n-1}\kappa_n\ra\la\kappa_n\kappa_1\ra}\,\Gamma(\Delta_1+1)\Gamma(\Delta_n+1)\prod_{j=2}^{n-1}\Gamma(\Delta_{j}-1)\times\\
    \int_{\mathbb{H}_3} \frac{\D^3x}{x^4}\,\frac{|x|^{\Delta_1+\Delta_n+2}(-\im)^{n-4-\Delta_1-\Delta_n-\sum_{k=2}^{n-1}\Delta_k}}{(x\cdot(\kappa_1\bar z_1))^{\Delta_1+1}(x\cdot(\kappa_n\bar z_n))^{\Delta_n+1}}\prod_{i=2}^{n-1}\frac{|x|^{\Delta_i-1}}{(x\cdot(\kappa_i\bar z_i))^{\Delta_i-1}}\,,
\end{multline}
where we recognize the final integral as the D-term that appears in contact Witten diagrams calculations. We note that similar results were obtained in \cite{Casali:2022fro} considering MHV amplitude in a certain chiral scalar coupled to Yang-Mills theory \cite{Dixon:2004za}. The reason \cite{Casali:2022fro} considered looking at Mellin transformed MHV amplitudes on $\mathbb{H}_3$ slices for a 4d Yang-Mills-scalar theory rather than 4d Yang-Mills theory is because of the distributional nature of the celestial amplitude of 4d conformally invariant theories. The addition of the scalar profile was necessary for computing the isolated integral transverse to the hyperbolic slices. Here by performing the scaling reduction on twistor space, we bypass the problem by recognizing the Mellin transformed wavefunctions are in fact minitwistor representatives on $\mathbb{MT}$. 

\subsection{Evaluating 3-pt}
By the $AdS_3/CFT_2$ dictionary, we can evaluate the D-term integral in \eqref{n_correlator} at 3 points \cite{Simmons_Duffin_2014}, which should give us hints on the form of the 2d CFT dual to our scaling reduced Yang-Mills theory. Recall the Mellin transform of exponentials 
\begin{equation}
    \frac{1}{\Gamma(m)}\int_{\mathbb{R}^*} \frac{\d s}{s} s^{m} e^{-bs} = \frac{1}{b^m}\,.
\end{equation}
Using this identity, we can rewrite the integrand in the D-term integral as
\begin{align}
        &\int_{\mathbb{H}_3} \frac{\D^3x}{x^4}\,\frac{|x|^{\Delta_1+\Delta_2+\Delta_3+1}(-\im)^{n-4-\Delta_1-\Delta_n-\sum_{k=2}^{n-1}\Delta_k}}{(x\cdot(\kappa_1\bar z_1))^{\Delta_1+1}(x\cdot(\kappa_2\bar z_2))^{\Delta_2+1}(x\cdot(\kappa_3\bar z_3))^{\Delta_3-1}} \nonumber\\
        = &\frac{1}{\Gamma(-\mathbf{\Delta})\Gamma(\Delta_1+1)\Gamma(\Delta_2+1)\Gamma(\Delta_3-1)}\int s^2\D^3 x \,\frac{\d s}{s}\bigwedge_{i=1}^3 \left(\frac{\d t_i}{t_i}\right) t_1^{\Delta_1+1}t_2^{\Delta_2+1}t_3^{\Delta_3-1}\,\e^{sx^2+\sum_{i=1}^3 t_i(x\cdot p_i)}\,,
\end{align}
where $\mathbf{\Delta}=\left(1+\sum_{k=1}^3\Delta_k\right)/2$ and we have rescaled $t\rightarrow st$ to cancel the unwanted powers of $s$ depending on $\Delta_i$s. Ignoring the Gamma functions for now for presentation purposes, we then change the integration variable $sx^2=x_{\text{4d}}^2$
\begin{multline}
    \int \d^4 x \bigwedge_{i=1}^3 \left(\frac{\d t_i}{t_i}\right)  t_1^{\Delta_1+1}t_2^{\Delta_2+1}t_3^{\Delta_3-1}\,\e^{x^2_{\text{4d}}+\sum_{i=1}^3 t_i(x_{\text{4d}}\cdot p_i)} \\= \int \d^4 x \bigwedge_{i=1}^3 \left(\frac{\d t_i}{t_i}\right) t_1^{\Delta_1+1}t_2^{\Delta_2+1}t_3^{\Delta_3-1}\,\e^{(x_{\text{4d}}-\frac{1}{2}\sum_{i=1}^3p_it_i)^2+\frac{1}{2}\sum_{i\neq j}p_i\cdot p_jt_it_j}\,,
\end{multline}
where we have completed the square on the exponential. The $x_{\text{4d}}$-integral is just a Gaussian integral now, evaluating it leaves us with 
\begin{equation}
    \int\bigwedge_{i=1}^3 \left(\frac{\d t_i}{t_i}\right) t_1^{\Delta_1+1}t_2^{\Delta_2+1}t_3^{\Delta_3-1}\,e^{\frac{1}{2}\left(p_1\cdot p_2t_1t_2+p_2\cdot p_3t_2t_3+p_3\cdot p_1t_1t_3\right)}\,.
\end{equation}
A further change of variables $s_3=t_1t_2,\,s_2=t_1t_3,\,s_1=t_2t_3$ allows us to disentangle the integration parameter on the exponential 
\begin{equation}
    \frac{1}{2}\,\int\bigwedge_{i=1}^3 \left(\frac{\d s_i}{s_i}\right) s_1^{\Delta_2+\Delta_3-\Delta_1-1}s_2^{\Delta_1+\Delta_3-\Delta_2-1}s_3^{\Delta_1+\Delta_2-\Delta_3+3}\,\e^{\frac{1}{2}\left(p_1\cdot p_2s_3+p_2\cdot p_3s_1+p_3\cdot p_1s_2\right)}\,.
\end{equation}
Putting back the Gamma function prefactors after evaluating all the integrals gives the 3-point function in the dual 2d CFT:
\begin{multline}
    \widetilde{\mathcal{M}}_{3} = \frac{1}{\Gamma(-\mathbf{\Delta})}\,\frac{\la 13\ra^4}{\la 12\ra\la 23\ra\la 31\ra}\,\\
    \frac{1}{(\la 12\ra[12])^{\Delta_1+\Delta_2-\Delta_3+3}(\la 23\ra[23])^{\Delta_2+\Delta_3-\Delta_1-1}(\la 31\ra[31])^{\Delta_1+\Delta_3-\Delta_2-1}}\,.\label{CFT_correlator}
\end{multline}
The D-term integral gives the canonical 3-particle conformal invariant, as expected. Note that there are dependence on $[\,,\,]$ in the correlator for generic $\Delta_i$s, which suggests that the boundary 2d CFT have independent holomorphic and antiholomorphic sectors. However the appearance of the Park-Taylor factor in front indicates that the correlator breaks conformal symmetry, suggesting the 2d CFT we are considering is actually some "chirally dressed" CFT, this was also observed in the three and four point calculations in \cite{Casali:2022fro}.

\section{Discussion}\label{discussion}
So far the discussions on celestial holography have mainly been centered around two different angles. In \cite{Costello:2020jbh}, the authors proposed to approach flat space holography using Koszul duality. This amounts to showing the existence of a certain 2d chiral algebra dual to a system on a defect inserted in the bulk of 6d $\mathbb{PT}$ coupling to the existing self-dual Yang-Mills theory. MHV form factors in the 4d bulk is then computed by inserting gauge invariant local operator $\tr(B^2)$, which can also be obtained from the correlation functions of the 2d chiral algebra generators on the boundary sphere. However, due to the absence of the momentum conserving delta function, this 2d chiral algebra only captures part of the dynamics in the bulk.

On the other hand, celestial amplitudes computed directly by Mellin transforming 4d Minkowski space amplitudes seem to have distributional delta functions $\bar\delta(z-\bar z)$ mixing the holomorphic and antiholomorphic sectors, which suggests certain gnarly properties of the dual CCFT. Namely the CCFT might not have a cleanly separated holomorphic and antiholomorphic sectors like in conventional 2d CFTs. 

The approach of scale reducing twistor actions to projective spinor bundles of hyperbolic 3-slices seems to avoid such issues, thanks to the natural interpretation of Mellin transform as a projective integral, which enables us to see 4d Minkowski space as the embedding space of $AdS_3$. 
In principle, other 4d conformal theories can also be manipulated in the way described in this paper. Exploiting the properties of minitwistor space should give us new insights into the structure of CCFT dual to these 4d theories from a twistor action perspective of $AdS_3/CFT_2$. 

The geometric structure of minitwistor space $\mathbb{MT}$ also makes it easy to insert non-local giant gravitons stretching along the geodesics of $AdS_3$ \cite{McGreevy:2000cw}, which localizes at a point on $\mathbb{MT}$. More recently, these objects were studied in the context of twisted holography \cite{Gaiotto:2021xce,Budzik:2021fyh}. 

Geometrically, we have regarded $\mathbb{R}^4$ as the embedding space of $\mathbb{H}_3$ in this paper, where Mellin transforms on momentum eigenstates in $\mathbb{R}^4$ can be seen as dilatation eigenstates on $\mathbb{H}_3$ as in the embedding formalism. Via the usual $AdS_3/CFT_2$ correspondence, we see flat holography in the time-like region as a byproduct of the embedding. One could always ask how robust such constructions are in generic dimensions. For example, starting with $AdS_{d+1}$, the relation between the $d+2$-dimensional embedding space of $AdS_{d+1}$ and its $d$-dimensional conformal boundary could perhaps provide an alternative perspective on celestial holography.

\section*{Acknowledgements}
It is a pleasure to thank Tim Adamo and David Skinner for discussions. We also thank Tim Adamo, Simon Heuveline and David Skinner for commenting on earlier versions of the draft. A large amount of the twistor theory (including all sections without citations to extant work) is due to many productive discussions and original work by SS with Yvonne Geyer, Lionel Mason and especially David Skinner about the corresponding story in $AdS_5$, minimally adapted to the $AdS_3$ case. WB is supported by the Royal Society Studentship. SS is supported by the Trinity College Internal Graduate Studentship.  

\newpage
\appendix

\section{Recap of embedding space formalism for $\mathbb{H}_{d+1}, \mathbb{R}^d$ in $\mathbb{R}^{d+1,1}/\mathbb{R}^*$}\label{Appendix_A}

Note that we will always work in mostly minus signature. Although the following sections specialise to the case of $\mathbb{R}^{d+1,1}$ for definiteness, any other signature can be described in the same way with minor substitutions. Good references include \cite{Simmons_Duffin_2014}, \cite{Weinberg_2010} and \cite{Costa_2014}.

\subsection{Embedding Formalism for $\mathbb{R}^d$}
In the embedding space description of a CFT living in $\mathbb{R}^d$, the CFT primaries are pulled back to the projectivized null cone of the origin (with the origin removed) $\{\mathbb{R}^{d+1,1}\slash \{(0,0,...,0)\}\}/\mathbb{R}^*$ in $\mathbb{R}^{d+1,1}$ embedding space and conventionally (we stress that there is no \emph{a priori} necessity) are assigned $d+2$-dimensional projective scaling weight to agree with their dilatation weight $\Delta$:
\begin{equation}
    X=\{(X_+,X_{-},\vec{X}_d)/ X \sim r X, r \in \mathbb{R}^*\} \in \mathbb{R}^{d+1,1}/\mathbb{R}^*, \,\,\,\, \vec{x}_d \in \mathbb{R}^d\,;
\end{equation}
\begin{equation}
    \phi_{\Delta}(\vec{x}_d) \rightarrow \Phi_{\Delta}(X), \,\,\,\, X^2 =  X_+ X_- - \vec{X}_d^2 = 0\,;
\end{equation}
\begin{equation}
    X \cdot \frac{\partial}{\partial X} \Phi_{\Delta}(X) = \Delta \Phi_{\Delta}(X)\,.
\end{equation}
The explicit relation between the coordinates $X$ and $\vec{x}_d$ is:
\begin{equation}
    X = X_+(1,x_d^2, \vec{x}_d) / X \sim r X, r \in \mathbb{R}^*, \,\,\,\, X^2 =X_+^2( 1 \times x_d^2 -x_d^2) = 0\,.
\end{equation}
Conformal invariant expressions in $\vec{x}_d$ become Lorentz invariant expressions in $X$. In particular, dilatation invariant expressions in $\vec{x}_d$, under this assignation of $d+2$-dimensional scale weight, correspond to scaling invariant expressions in $X$. 

Notice that the projective scaling of $X$ is a gauge freedom that has to be fixed, which is usually done by demanding that the first entry, $X^+$, be 1. Having done so, the nonlinear action of the conformal group on $\vec{x}_d$ can be resolved into the linear action of the $d+2$ dimensional Lorentz group on $X$ and a gauge transformation to return to the chosen gauge slice.

Notice that we are reading $X$ as homogeneous coordinates on the projective space $\{\mathbb{R}^{d+1,1}\slash \{(0,0,...,0)\}\}/\mathbb{R}^*$, and $(1,x_d^2,\vec{x}_d)$ as inhomogeneous coordinates on the patch where $X_+ \neq 0$. Since we are only concerned with the null cone of the origin (with the origin removed), we are always working on that patch.

\subsection{Embedding formalism for $\mathbb{H}_{d+1}$}
The projective null cone of the origin can be understood as the boundary of the projective causal future of the origin of embedding space. From this perspective, it is natural that a choice of metric on the causal future of the origin of embedding space will allow us to describe the $\mathbb{H}_{d+1}$ whose conformal boundary is the $\mathbb{R}^d$ which has been identified as the projective null cone of the origin. More explicitly, let us restrict to the causal future of the origin:
\begin{equation}
    X^2 = (X_+,X_{-},\vec{X}_d)^2 = X_+ X_- - \vec{X}_d^2 > 0 \,.
\end{equation}
Now consider the following (basic and invariant under the Euler vector field $X\cdot \partial_X $) metric:
\begin{equation}
    ds^2_{\mathbb{H}} = \frac{1}{X^2}\left(dX^{\mu}-\frac{X^\mu (X \cdot dX)}{X^2}\right)^2 = \frac{1}{X^2}\left(dX^2+(-2+1)\frac{(X\cdot dX)^2}{X^2}\right)\,.
\end{equation}
It is constructed from the flat embedding space metric by projecting out the components of $dX^{\mu}$ that point in the scaling direction using the projection matrix (so that it is basic under the Euler vector field $i_{X\cdot \partial_X}g $):
\begin{equation}
    P^{\mu}_{\nu} = \delta^\mu_\nu - \frac{X^\mu X_\nu}{X^2}\,,
\end{equation}
and it is divided by $X^2$ such that it is invariant under the projective scaling (equivalently, so that it is invariant under the Euler vector field: $\mathcal{L}_{X \cdot \partial_X} g = 0$). To make contact with the upper half plane coordinates in hyperbolic space $x_{d+1}=(x_0, \vec{x}_d)$, work in the patch where $X_+ \neq 0$:
\begin{equation}
    X = X_+(1, x_0^2+x_d^2, \vec{x}_d)/X \sim r X, r \in \mathbb{R}^*\,.
\end{equation}
The coordinates $(x_0^2+x_d^2, \vec{x}_d)$ are inhomogeneous coordinates on the patch of the projectivized embedding space where $X_+ \neq 0$. Notice that the action of $SO(d+1,1)$ is realised nonlinearly on the $(x_0,\vec{x}_d)$ because the linear action on $X$ takes us off the gauge slice where the first entry of $X$ is 1, and we have to return to the gauge slice with a scaling. In these coordinates, the given metric is:
\begin{equation}
    ds^2_{\mathbb{H}} = \frac{1}{X^2}\left(dX^2-\frac{(X\cdot dX)^2}{X^2} \right) = \frac{dx_0^2+d\vec{x}_d^2}{x_0^2}\,,
\end{equation}
which is our familiar upper half plane metric for hyperbolic space. In a similar vein, the condition that we are in the causal future of the origin is expressed as:
\begin{equation}
    X^2 = X_+^2 \left(x_0^2+\vec{x}_d^2-\vec{x}_d^2\right) = X_+^2 x_0^2 > 0\,.
\end{equation}

Notice that the conventional description of a hyperboloid as $X^2=R^2$ for some constant $R$ can be recovered as a gauge fixing for the above equation, where the arbitrary scale of $X$ is fixed such that we have $X^2=R^2$. The advantage to working in this projective formalism is that we are better suited to make contact with the embedding space description of a putative dual theory on the conformal boundary.


\section{Recap of needed twistor theory}\label{Appendix_B}
Here we provide a swift recap of the required twistor theory. Standard references include \cite{Bailey_Dunne_1998,Huggett_Tod_1985,Penrose:1986ca}. A good discussion of $\mathbb{PS}$ for the case of reduction to $\mathbb{R}^3$ can be found in \cite{Adamo:2017xaf}. A good discussion of minitwistors and minitwistor methods for the corresponding story in 5D can be found in \cite{Adamo:2016rtr}.

\subsection{Recap of Minitwistor theory}\label{appendix_minitwistor}
\subsubsection{The projective spin bundle $\mathbb{PS}$}\label{section_PS}
In our discussion of embedding space we have established that 3D spheres / pseudospheres / hyperboloids can be identified as (a submanifold of) the projectivzation of the 4D embedding space of some signature. There is an analogous story for twistor space, which holds in all signatures but is most well-motivated by considering Euclidean signature embedding space.

Consider flat 4D twistor space $\mathbb{PT}, I_{AB} = \textrm{diag}(0_2,\epsilon^{\alpha \beta})$
\begin{equation}
    \mathbb{PT} = \mathbb{CP}^3 \setminus \mathbb{CP}^1_I = \{Z^A = (\lambda_{\alpha}, \mu^{\dot \alpha}) \in \mathbb{CP}^3 | \epsilon_{ABCD}Z^CZ^D \neq I_{AB} \iff \lambda^{\alpha} \neq 0\}\,.
\end{equation}
Each point in twistor space corresponds to a totally null 2-plane in $\mathbb{C}^4$. The intersection of this null 2-plane with the real Euclidean slice of $\mathbb{R}^4 \in \mathbb{C}^4$ must be dimension 0 as the tangent spaces are transverse. By linearity, the intersection is either empty or a single point. We can explicitly solve for the point and show that the intersection is nonempty:
\begin{equation}
   \mu^{\dot \alpha}= x^{\alpha \dot \alpha} \lambda_{\alpha} \quad \textrm{and} \quad x^{\alpha \dot \alpha} = \hat x^{\alpha \dot \alpha} \implies x^{\alpha \dot \alpha} = \frac{\hat \mu^{\dot \alpha}\lambda^{\alpha}-\mu^{\dot \alpha}\hat \lambda^{\alpha}}{\langle \lambda \hat \lambda \rangle}\,.
\end{equation}
Therefore there exists a projection map $\pi: \mathbb{PT} \rightarrow \mathbb{R}^4, Z^A \rightarrow x^{\alpha \dot \alpha}$ with a 2-real-dimensional kernel, and it is natural to consider $\mathbb{PT}$ as the total space of a fibre bundle over $\mathbb{R}^4$ with a 2-real-dimensional fibre. The conventional description is what is known as the projective spin bundle, $\mathbb{PS}_4$, coordinatized by $(\lambda_{\beta}, x^{\alpha \dot \alpha}) \in \mathbb{PS}_4 = \mathbb{CP}^1\times\mathbb{R}^4  $, the trivial bundle of projectivized undotted Weyl spinors over $\mathbb{R}^4$. As a real manifold, $\mathbb{PS}_4 \cong \mathbb{PT}$, with $(\lambda_{\alpha},x^{\alpha \dot \alpha}\lambda_{\alpha})=(\lambda_{\alpha}, \mu^{\dot \alpha})$. Notice that the $\mathbb{CP}_I^1$ we removed from $\mathbb{CP}^3$ to construct $\mathbb{PT}$ corresponds to the point $x =\infty$ and the 0 spinor, both of which are not present in $\mathbb{PS}_4$.

With this identification between $\mathbb{PS}_4$ and $\mathbb{PT}$ as real manifolds, we can read off what a real scaling reduction on the spacetime does to the twistor space. Taking the real scaling reduction on $\mathbb{PS}_4$, we get $\mathbb{RP}^3 \times \mathbb{CP}^1$, the trivial bundle of projectivized undotted 4D Weyl spinors over $\mathbb{RP}^3$, which we will denote $\mathbb{PS}$. As discussed before (in various signatures), we can pick a metric on $\mathbb{RP}^3$ such that it is $\mathbb{H}_3$. With this choice of metric, we preserve the $SO(4)$ isometries of the original spacetime and the spin structure is identical. Therefore $\mathbb{PS}$ is still the trivial bundle of undotted Weyl spinors over the scaling reduced real Euclidean spacetime.

If we do not choose to work in these coordinates, we have the generic description of $\mathbb{PS}$:
\begin{equation}
    \mathbb{PS} = \{(\lambda_{\alpha},\mu^{\dot\alpha}) \in \mathbb{PT} | \mu^{\dot\alpha} \sim r \mu^{\dot\alpha}, \quad r \in \mathbb{R}^*\}\,.
\end{equation}
With incidence relation:
\begin{equation}
    \mu^{\dot\alpha} = x^{\alpha \dot\alpha} \lambda_{\alpha} \,,
\end{equation}
with equivalence up to real scaling. There are pros and cons to working with $\mathbb{PS}$. Firstly, the space fibres over the spacetime, and therefore it is easier to see what objects on $\mathbb{PS}$ correspond to on the spacetime either by integrating out the fibres or projecting out components of forms. However, $\mathbb{PS}$ is a 5-real-dimensional CR manifold rather than an honest complex manifold, and slightly awkward to work with. The other possible space that originates from a scaling reduction of $\mathbb{PT}$ is a complex scaling reduction of $\mathbb{PT}$, which can be shown to be a complex manifold. We will show that it is in fact the minitwistor space $\mathbb{MT}$ of the complexification of hyperbolic 3-space $\mathbb{H}_3$.

\subsubsection{The Minitwistor space $\mathbb{MT}$ of $\mathbb{H}_3$}
The minitwistor space $\mathbb{MT}$ of the complexification of $\mathbb{H}_3$ (equivalently $\mathbb{H}_3$ / $dS_3$ / $S^3$ with minor substitutions) is the space of oriented geodesics or equivalently, ordered pairs of points on the complexified conformal boundary $S^{2}$. We claim and demonstrate that:
\begin{enumerate}
  \item $\mathbb{MT}$ arises from a complex scaling reduction of $\mathbb{PT}$ (resp. a $U(1)$ reduction of $\mathbb{PS}$)
  \item $\mathbb{MT} = \{ \mathbb{CP}^1 \times \mathbb{CP}^1 \} \setminus \mathbb{CP}^1$
  \item $\mathbb{MT}$ is the bitwistor space of (conformally) flat 2D with the diagonal removed.
\end{enumerate}
Of the real slices of $\mathbb{H}_{3 \mathbb{C}}$, the reason we choose to describe $\mathbb{H}_{3}$ in this section is because the conformal boundary is real Euclidean $S^2$, which will prove to be more convenient to describe than other real slices of $S^2_{\mathbb{C}}$. Therefore we now consider real Lorentzian embedding space $x^{\alpha \dot \alpha} = \bar x^{\alpha \dot \alpha}$, with the barred Lorentzian conjugation:
\begin{equation}
    \bar \mu^{\alpha} = \begin{pmatrix}
        \bar \mu^{\dot 0} \\ \bar \mu^{\dot 1}
    \end{pmatrix} \quad,\quad \bar \lambda_{\dot \alpha} = \begin{pmatrix}
        \bar \lambda_{0} \\ \bar \lambda_{1}
    \end{pmatrix}\,.
\end{equation}
Consider the complex scaling reduction of $\mathbb{PT}$ defined by taking our above prescription for landing on $\mathbb{PS}$ and promoting the scaling from a $\mathbb{R}^*$ scaling to a $\mathbb{C}^*$.
\begin{equation}
    \mathbb{MT}' = \{(\lambda_{\alpha}, \mu^{\dot\alpha}) \in \mathbb{PT} | \mu^{\dot\alpha} \sim r \mu^{\dot\alpha}, \quad r \in \mathbb{C}^*\} = \{(\lambda_{\alpha},\mu^{\dot\alpha}) \in \mathbb{CP}^1_{\lambda} \times \mathbb{CP}^1_{\mu} \}\,.
\end{equation}
In the second equality, we have played slightly fast and loose - while $\mu^{\dot \alpha}$ is strictly speaking a $\mathcal{O}(1) \oplus \mathcal{O}(1)$ bundle over $\mathbb{CP}^1$ we have treated it as if it were a $\mathbb{C}^2$ and performed the reduction, identifying the resulting space as $\mathbb{CP}^1$. The equality can be seen as a motivation for this definition of $\mathbb{MT}'$ rather than a true equality. A point in $\mathbb{MT}'$ corresponds to an (ordered, as we can distinguish $\mu$ from $\lambda$) pair of real Lorentzian points $(z^{\alpha \dot\alpha},w^{\alpha \dot\alpha})$ on the boundary $S^2$, written in embedding space coordinates as:
\begin{equation}
    z^{\alpha \dot\alpha} = \bar \mu^{\alpha} \mu^{\dot\alpha} \quad,\quad w^{\alpha \dot\alpha} = \lambda^{\alpha} \bar \lambda^{\dot \alpha}\,,
\end{equation}
that define an oriented (as we can distinguish $z$ from $w$) geodesic through the bulk $\mathbb{H}_3$:
\begin{equation}
    x^{\alpha \dot\alpha}(s) = z^{\alpha \dot\alpha} e^s + w^{\alpha \dot\alpha} e^{-s} \quad , \quad x^{\alpha \dot\alpha}(s) \lambda_{\alpha} = \mu^{\dot\alpha}\,,
\end{equation}
$z$ and $w$ are distinct boundary points except when $\bar \mu^{\alpha} \propto \lambda^{\alpha} \iff \langle \bar \mu \lambda \rangle = 0$. The locus of this equation is (the graph of a) $\mathbb{CP}^1$ in $\mathbb{MT}'$, and must be removed. It is known as the antiholomorphic diagonal, or $\mathbb{B}_{E2}$, the real Euclidean bitwistor space of the boundary $S^2$, which can be shown to be equivalent to a single copy of the boundary $S^2$ as a real manifold. Therefore we have $\mathbb{MT}$ defined as:
\begin{equation}
    \mathbb{MT} = \mathbb{MT}' \setminus \mathbb{B}_{E2} = \{(\lambda_{\alpha},\mu^{\dot\alpha}) \in \mathbb{CP}^1_{\lambda} \times \mathbb{CP}^1_{\mu} | \langle \bar \mu \lambda \rangle \neq 0\}\,.
\end{equation}
Note that the removal of this $\mathbb{CP}^1$ is crucial to getting nontrivial cohomology on $\mathbb{MT}$. Discussion of the twistor space, (am)bitwistor space, and real Euclidean bitwistor space of flat 2D can be found in the next appendix for the curious reader, as their inclusion here is too much of a diversion. We assemble some correspondences between points and lines in $\mathbb{MT}$ and $\mathbb{H}_{3\mathbb{C}}$, the associated spacetime.
\begin{enumerate}
  \item As demonstrated, a point in $\mathbb{MT}$ corresponds to an oriented geodesic on the spacetime
  \item A point $x$ in $\mathbb{H}_{3\mathbb{C}}$ corresponds to a line (graph of a $\mathbb{CP}^1$) $L_x$
  \item If lines $L_x, L_y$ in $\mathbb{MT}$ intersect in exactly one point, this corresponds to null separation of $x,y$ on $\mathbb{H}_{3\mathbb{C}}$
\end{enumerate}
To demonstrate one direction of the correspondence in point 2, consider a point $x$. Then the line $L_x$ it defines on $\mathbb{MT}$ is $(\lambda_{\alpha}, x^{\alpha \dot\alpha} \lambda_{\alpha})$. For the other direction, consider the line (holomorphically embedded $\mathbb{CP}^1$) defined by minitwistor points $(\lambda_{1\, \alpha}, \mu^{\dot\alpha}_1),(\lambda_{2\, \alpha}, \mu^{\dot\alpha}_2)$. The fact that it is a linear embedding means that there exists some matrix $m^{\alpha \dot \alpha}$ defined up to complex scaling that satisfies:
\begin{equation}
    \mu^{\dot\alpha}_{i} = m^{\alpha \dot \alpha} \lambda_{i\, \alpha} \quad , \quad i=1,2\,.
\end{equation}
Each equation determines $m^{\alpha \dot \alpha}$ up to adding terms proportional to $\lambda_{i \, \alpha}$, and therefore for non-degenerate ($\langle \lambda_1 \lambda_2 \rangle \neq 0$ and $[\mu_1 \mu_2]\neq 0$) lines in $\mathbb{MT}$, this uniquely defines a spacetime point $m^{\alpha \dot \alpha}$.

For point 3, consider two minitwistor lines $L_x$ and $L_y$ that intersect in $\mathbb{MT}$ at $(\mu^{\dot\alpha},\lambda_{\alpha})$:
\begin{equation}
    \mu^{\dot\alpha} = x^{\alpha \dot \alpha} \lambda_{\alpha} \quad,\quad \mu^{\dot\alpha} = y^{\alpha \dot \alpha} \lambda_{\alpha} \quad \implies x^{\alpha \dot\alpha} y_{\dot\alpha}^{\beta} \lambda_{\alpha}\lambda_{\beta} = 0\,.
\end{equation}
This equation is quadratic in $\lambda$ and generically has two solutions. If we claim that the lines intersect in exactly one point, we impose that there is a double zero, $x^{\dot\alpha (\alpha} y_{\dot\alpha}^{\beta)} = |x||y|o^{\alpha}o^{\beta}$ for some constant spinor $o$. Then we have:
\begin{equation}
    x^{\dot\alpha \alpha} y_{\dot\alpha}^{\beta} = |x||y|o^{\alpha}o^{\beta} + \frac{1}{2} \epsilon ^{\alpha \beta}x \cdot y \implies x^2 y^2 = (x\cdot y)^2\,.
\end{equation}
Recall that the geodesic distance $d(x,y)$ on $\mathbb{H}_3$ (with signs and factors of $i$ for other signatures) satisfies: 
\begin{align}
    \cosh (d(x,y)) = {\frac{x\cdot y}{|x||y|}}\,.
\end{align}
Hence if $L_x$ and $L_y$ intersect at one point, $\cosh(d(x,y)) = 1$ and $x,y$ are null separated. Therefore we see that $\mathbb{MT}$ encodes the conformal structure of $\mathbb{H}_{3 \mathbb{C}}$ in much the same way as $\mathbb{PT}$ encodes the conformal structure of conformally flat 4D.

\subsubsection{Cohomology classes and the Penrose transform on $\mathbb{MT}$}
Just as in flat space, there are correspondences between $H^1$s of various weights on $\mathbb{MT}$ and spacetime fields on $\mathbb{H}^3$ \cite{Bailey_Dunne_1998}, inherited from the familiar correspondences between $H^1$s on $\mathbb{PT}$ and flat 4D \cite{Huggett_Tod_1985}\cite{Penrose:1986ca}. There are two cases, colloquially known the the "positive" and "negative" weight cases, where the names require some explanation, and both descend to statements on $\mathbb{MT}$ with minor modifications.

To set the stage, consider the representative of a Dolbeault cohomology class $f \in H^{0,1}(\mathbb{PT},\mathcal{O}(p-2))$. For $p\leq 0$, this is known as the negative weight case, and the correspondence is the Penrose transform onto on-shell fields (all with totally symmetrized indices):
\begin{align}
    \{f_{p-2}\in H^{0,1}(\mathbb{PT},\mathcal{O}(p-2))\} \cong \{\ \phi_{\alpha_1 ... \alpha_p}(x) | \square \phi_{\alpha_1 ... \alpha_p}(x) = 0=\partial^{\alpha_1 \dot\alpha} \phi_{\alpha_1 ... \alpha_p}(x)\}\,.
\end{align}
For $p>0$, this is known as the positive weight case, and the correspondence is the Sparling transform onto potentials modulo gauge (all with totally symmetrized indices):
\begin{align}
    \{f_{p-2}\in H^{0,1}(\mathbb{PT},\mathcal{O}(p-2))\} \cong \{\ A^{\dot\alpha}_{\alpha_1 ... \alpha_p}(x) | A^{\dot\alpha}_{\alpha_1 ... \alpha_p}(x) \sim A^{\dot\alpha}_{\alpha_1 ... \alpha_p}(x) + \partial^{\dot \alpha}_{(\alpha_1}g_{\alpha_2 ... \alpha_p)}(x)\}\,.
\end{align}
For both of these, the correspondence one way is realised by integral formulae, and the properties of the spacetime fields can be verified by direct substitution. For $p\leq 0$:
\begin{equation}
    \phi_{\alpha_1 ... \alpha_p}(x) = \int_{\mu=x\lambda} \D \lambda \lambda_{\alpha_1}...\lambda_{\alpha_p} f\,.
\end{equation}
For $p > 0$ (this integral transform is proven to be equivalent to the standard argument due to Sparling in the appendix):
\begin{equation}
    A^{\dot\alpha}_{\alpha_1 ... \alpha_p}(x) = \int_{\mu=x\lambda} \D \lambda \prod \frac{v_{i\, \alpha_i}}{\la \lambda v_{i} \ra} \frac{\partial}{\partial \mu_{\dot \alpha}} f\,.
\end{equation}
The point of working in the embedding formalism for spacetime is that we can import these results directly to $\mathbb{MT}$ without much fuss. $H^{0,1}$ on $\mathbb{MT}$ can be pulled back to $H^{0,1}$ on $\mathbb{PT}$, corresponding to spacetime fields on embedding space. These are then interpreted as embedding space representatives of fields on $\mathbb{H}_3$. Suitably interpreted, this lets us construct correspondences between $H^{0,1}$ on $\mathbb{MT}$ and spacetime fields on $\mathbb{H}_3$.

First notice that the integral formulae above realise the correspondence in exactly the same way as in the flat 4D case. The outputs come out with scaling weight, indicating that they do not correspond directly to fields on $\mathbb{H}_3$ but rather to the conventional embedding space representation of fields of defined dilatation weight on $\mathbb{H}_3$, where we have multiplied in a prefactor $|x|^{\Delta}$ to trade the dilatation weight for scaling weight. In order to land on the fields on $\mathbb{H}_3$ directly, we must use the scaling weight-free incidence relations $\mu = \frac{x}{|x|} \lambda$, or equivalently multiply in the appropriate power of $|x|$.

\subsection{Twistors in (conformally) flat 2D}
\subsubsection{The twistor space}
The projective twistor space for compactified 2 dimensional (conformally) flat spacetime is $\mathbb{CP}^1$. We may determine this from the usual definition of twistors as follows. A twistor for flat $\mathbb{R}^{p,q}$ transforms in the fundamental of the spin representation of $SO(p+1,q+1)$, which is colloquially said as "Twistors are spinors for 2D higher". This is the same as the statement that twistors transform linearly under the action of the conformal group. In the case of $d=2$, the most convenient thing to do is consider $\mathbb{R}^2$, and therefore consider proper orthochronous $SO(1,3) \cong SL(2)_{\mathbb{C}}/\mathbb{Z}_2$. From the commutation relations, we can identify the generators $P^{\mu},K_{\mu},L,D$ as follows, formed from complex linear combinations of the $SL(2)$ generators:
\begin{align}
    iP^1&=P^2:=\begin{pmatrix} 0 & i \\ 0 & 0\end{pmatrix},\,\,\,\,iK_1=K_2:=\begin{pmatrix} 0 & 0 \\ i & 0\end{pmatrix}
    \\
    L&:=L^{[12]}=\frac{1}{2}\begin{pmatrix} i & 0 \\ 0 & -i\end{pmatrix},\,\,\,\,D:=\frac{1}{2}\begin{pmatrix} 1 & 0 \\ 0 & -1\end{pmatrix}\,.
\end{align}
Under this identification, we can read off that the action of $P$ on a twistor $\lambda^{\alpha}=(\lambda^0, \lambda^1) \in \mathbb{C}^2$ is to translate $\lambda^0$ by some amount of $\lambda^1$:
\begin{equation}
    (\epsilon_1 P^1 + \epsilon_2 P^2)(\lambda^0,\lambda^1)=(\lambda^0+(\epsilon_1 + i\epsilon_2)\lambda^1,\lambda^1)\,.
\end{equation}
Consider $Z=x+\im y, \, (x,y) \in \mathbb{R}^2$ with the natural translation action on $\mathbb{R}^2$:
\begin{equation}
    (\epsilon_1 P^1 + \epsilon_2 P^2)(x+\im y)=(x+\epsilon_1)+\im(y+\epsilon_2)\,.
\end{equation}
Since the incidence relations must be covariant under the action of the conformal generators, we may write the covariant incidence relation:
\begin{equation}
    x + \im y = \frac{\lambda_0}{\lambda_1}\,.
\end{equation}
We can see that this incidence relation uniquely specifies a point in $\mathbb{R}^2$. This is analogous to the statement that for flat 4D, the twistor space fibres over the spacetime because the null planes only intersect the Euclidean real slice at a single point. For complexified spacetime, the incidence relation determines a complex null line, which in $(1,1)$ signature ($\bar x = x, \bar y = -y$) are the right-moving geodesics:
\begin{equation}
    \delta x+\im\delta y = 0 \implies \delta x^2+\delta y^2 = 0\,.
\end{equation}
Note that the region of twistor space corresponding to the point at infinity for compactified $\mathbb{R}^2$ is $\lambda^1=0$, which can be deduced because it is preserved by the translation generators, and therefore we should remove it when considering the twistor space of $\mathbb{R}^2$. Having done so, we may work in inhomogeneous coordinates on the patch of the projective twistor space where $\lambda_1 \neq 0$ and we find that a point $(z,1) \in \mathbb{PT}_{\mathbb{R}^2}$ obeys $z=x+iy$, $(x,y) \in \mathbb{R}^2$. The twistor space is therefore just the spacetime equipped with the canonical complex structure $\mathbb{R}^2 \cong \mathbb{C}$. Therefore we have the somewhat tautological 2D Penrose transform:
\begin{equation}
    \{ H^{0,0}(\mathbb{PT}_{\mathbb{R}_2},\mathcal{O})\} \cong \{ \phi(x,y) | (x,y) \in \mathbb{R}^2, \,\, \bar \partial \phi = 0\}\,,
\end{equation}
where the proof is left as an exercise for the reader. We can also consider the barred twistor space coordinatized by $\mu_{\dot \alpha} \in \bar{\mathbb{PT}}_2$. For $\la \lambda \bar \mu \ra$ to be an $SL(2)$ scalar, we see that:
\begin{equation}
    (\epsilon_1 P^1 + \epsilon_2 P^2)(\mu_{\dot 0}, \mu_{\dot 1})=(\mu^{\dot 0}+(\epsilon_1 - \im\epsilon_2)\mu^{\dot 1},\mu^{\dot 1})\,.
\end{equation}
Such that the covariant incidence relation for the barred twistor space is:
\begin{equation}
    x - \im y = \frac{\mu_{\dot 0}}{\mu_{\dot 1}}\,.
\end{equation}
These incidence relations also determine a complex null line:
\begin{equation}
    \delta x-\im\delta y = 0 \implies \delta x^2+\delta y^2 = 0\,,
\end{equation}
where in $(1,1)$ signature ($\bar x = x, \bar y = -y$) we see that these are the left-moving null geodesics.

\subsubsection{The bitwistor space}
The ambitwistor space $\mathbb{A}_M$ of a spacetime $M$ is the space of null geodesics in the spacetime. In this case, it is $\bar{\mathbb{PT}}_2 \cup \mathbb{PT}_2$, which is not so interesting. Since 2D is so degenerate, we will look at something slightly different. Define the bitwistor space $\mathbb{B}_2$ to be:
\begin{equation}
    \mathbb{B}_2 = \{ (\lambda_{\alpha},\mu^{\dot \alpha}) \in \mathbb{PT}_2 \times \bar{\mathbb{PT}}_2 \}
\end{equation}
And define the Euclidean bitwistor space to be the antiholomorphic diagonal $\mathbb{CP}^1$:
\begin{equation}
    \mathbb{B}_{2E} = \{ (\lambda_{\alpha},\mu^{\dot \alpha}) \in \mathbb{PT}_2 \times \bar{\mathbb{PT}}_2| \la \bar \mu \lambda \ra = 0 \}
\end{equation}
The Euclidean bitwistor space is a $\mathbb{CP}^1$. As a real manifold it is equivalent to the Euclidean $S^2$.

\subsection{4D Potentials Penrose Transform}
\label{appendix_sparling_transform}
We demonstrate the equivalence of the integral transform given in the text with the standard argument (the argument due to Sparling) in the case of interest to us. The general weight case follows with minor substitutions.
\subsubsection{Integral Representation}
In this section we work in 4D and with an Abelian gauge group. To Penrose transform onto a z.r.m field with appropriate indices (symmetrized pair of dotted or undotted indices) to be a (anti) self-dual field strength $F_{\alpha \beta}$, we see that we start with a cohomology class $a \in H^{0,1}(\mathbb{PT},O)$ and perform
\begin{equation}
    F_{\dot \alpha \dot \beta}(x) = \int_{\mu = x \lambda} \langle \lambda d\lambda \rangle \frac{\partial}{\partial \mu^{\dot\alpha}}\frac{\partial}{\partial \mu^{\dot\beta}} a\,.
\end{equation}
The claim is that the integral representation of the Sparling transform to get (a representative of the equivalence classes of) the gauge field is as follows:
\begin{equation}
    A_{\alpha \dot \alpha}(x, v) = \int_{\mu = x \lambda} \langle \lambda d\lambda \rangle \frac{v_{\alpha}}{\langle \lambda v \rangle}\frac{\partial}{\partial \mu^{\dot\alpha}} a\,.
\end{equation}
We can see that this has the right weights for the projective integral to be well defined. To check that this does satisfy $F_{\dot\alpha \dot\beta} = \partial_{\dot\alpha}^{\alpha} A_{\dot\beta \alpha}$ we may plug the expressions in and verify. 
\begin{align}
    \partial_{\alpha}^{\dot\alpha} A_{\dot\beta \alpha} &= \int_{\mu = x \lambda} \langle \lambda d\lambda \rangle \frac{v_{\alpha}}{\langle \lambda v \rangle} \lambda^{\alpha} \frac{\partial}{\partial \mu^{\dot\alpha}} \frac{\partial}{\partial \mu^{\dot\beta}} a = F_{\dot\alpha \dot\beta}\,.
\end{align}
We see that the freedom to pick $v_{\alpha}$ is a gauge transformation:
\begin{align}
    A_{\alpha \dot \alpha}(x, v) -  A_{\alpha \dot \alpha}(x, u) &= \int_{\mu = x \lambda} \langle \lambda d\lambda \rangle \left( \frac{v_{\alpha}}{\langle \lambda v \rangle} - \frac{u_{\alpha}}{\langle \lambda u \rangle} \right) \frac{\partial}{\partial \mu^{\dot\alpha}} a \\
    &=
    \int_{\mu = x \lambda} \langle \lambda d\lambda \rangle \frac{\langle v u \rangle}{\langle \lambda v \rangle \langle \lambda u \rangle} \lambda_{\alpha} \frac{\partial}{\partial \mu^{\dot\alpha}} a \\
    &=\partial_{\alpha \dot \alpha }\int_{\mu = x \lambda} \frac{\langle \lambda d\lambda \rangle \langle v u \rangle}{\langle \lambda v \rangle \langle \lambda u \rangle} a \,,
\end{align}
where the second equality is easiest seen by contracting in an arbitrary dotted spinor and using the Schouten identity. Furthermore, despite the pole in the integrand, the prescription respects the cohomological freedom of $a$ as cohomology transformations are mapped to gauge transformations, which can be seen by an integration by parts:
\begin{align}
    \int_{\mu = x \lambda} \langle \lambda d\lambda \rangle \frac{v_{\alpha}}{\langle \lambda v \rangle}\frac{\partial}{\partial \mu^{\dot\alpha}} \bar \partial \phi = -(\lambda_{\alpha}\frac{\partial}{\partial \mu^{\dot\alpha}} \phi)_{\lambda = v} = \partial_{\alpha \dot \alpha} \phi_{\mu = x v, \lambda 
 =v}\,.
\end{align}
Note that this is also true for $v = \hat \lambda$, which is often a convenient choice:
\begin{align}
    \int_{\mu = x \lambda} \langle \lambda d\lambda \rangle \frac{\hat \lambda_{\alpha}}{\langle \lambda \hat \lambda \rangle}\frac{\partial}{\partial \mu^{\dot\alpha}} \bar \partial \phi = \int_{\mu = x \lambda} \frac{\D\lambda \D\hat \lambda}{\la \lambda \hat \lambda \ra^2} (\lambda_{\alpha}\frac{\partial}{\partial \mu^{\dot\alpha}} \phi) = \partial_{\alpha \dot \alpha}\int \frac{\D\lambda \D\hat \lambda}{\la \lambda \hat \lambda \ra^2}\phi (x\lambda,\lambda)\,.
\end{align}
\subsubsection{Proof}
The proof is a special case of the standard argument for the Sparling transform. As usual, we observe that $H^{0,1}(\mathbb{CP}^1)$ of weight 1 and below vanish and therefore the restriction of $a$ onto the $\mathbb{CP}^1_X$ (in $\mathbb{PT}$, defined by the incidence relations) is $\bar \partial$ exact:
\begin{equation}
    a|_{\mu = x \lambda} = \bar \partial h\,.
\end{equation}
In this case this relation is invertible, and up to some numerical factors we have
\begin{equation}
    h(x,C) - h(x,D) = \int \frac{\langle \lambda d \lambda \rangle \langle C D \rangle }{\langle C \lambda \rangle \langle \lambda D \rangle} a|_{\mu = x \lambda}\,.
\end{equation}
Pulled back to a line on twistor space, we see that we have the condition
\begin{equation}
    \bar \partial|_{\mu = x C} (C^{\alpha} \partial_{\alpha \dot \alpha} h) = \lambda^{\alpha} \partial_{\alpha \dot \alpha} a|_{\mu = x C} = 0\,,
\end{equation}
and therefore that $C^{\alpha} \partial_{\alpha \dot \alpha} h$ is a holomorphic function of $C$, and weight +1. Therefore it must be linear in $C$. Defining $C^{\alpha} A_{\alpha \dot \alpha}(x) := C^{\alpha} \partial_{\alpha \dot \alpha} h$, we have
\begin{align}
    A_{\alpha \dot \alpha}(x) &= \frac{\partial}{\partial C^{\alpha}} C^{\beta} \partial_{\dot\alpha \beta} h(x,C) \\
     &= \frac{\partial}{\partial C^{\alpha}} C^{\beta} \partial_{\dot\alpha \beta} \left(\int \frac{\langle \lambda d \lambda \rangle \langle C D \rangle }{\langle C \lambda \rangle \langle \lambda D \rangle} a|_{\mu = x \lambda} + h(x,D) \right) \\
    &=\left(\frac{\partial}{\partial C^{\alpha}} \int \frac{\langle \lambda d \lambda \rangle \langle C D \rangle }{\langle C \lambda \rangle \langle \lambda D \rangle} C^{\beta} \partial_{\dot \alpha \beta} a|_{\mu = x \lambda}\right) + \partial_{\alpha \dot \alpha} h(x,D)\\
     &=\left( \frac{\partial}{\partial C^{\alpha}} \int \frac{\langle \lambda d \lambda \rangle \langle C D \rangle }{\langle C \lambda \rangle \langle \lambda D \rangle} \langle C \lambda \rangle \frac{\partial}{\partial \mu^{\dot \alpha}} a|_{\mu = x \lambda} \right) + \partial_{\alpha \dot \alpha} h(x,D) \\
     &= \left( \int \frac{\langle \lambda d \lambda \rangle D_{\alpha}}{\langle \lambda D \rangle}  \frac{\partial}{\partial \mu^{\dot \alpha}} a|_{\mu = x \lambda} \right) + \partial_{\alpha \dot \alpha} h(x,D)\,.
\end{align}
We can drop the final term as a gauge transformation and recover the form as claimed.

\newpage
\bibliographystyle{JHEP}
\bibliography{ads3}

\end{document}